\documentclass[12pt]{article}
\usepackage{multirow}
\usepackage{epsfig}
\usepackage{amsmath}
\usepackage{hhline}
\usepackage{amssymb}
\usepackage{times}
\usepackage{cite}
\usepackage{dcolumn}

\newlength{\dinwidth}
\newlength{\dinmargin}
\begin{document}  
\newcommand{\beq}{\begin{displaymath}}
\newcommand{\eeq}{\end{displaymath}}

\newcommand{\pom}{{I\!\!P}}
\newcommand{\reg}{{I\!\!R}}
\newcommand{\slowpi}{\pi_{\mathit{slow}}}
\newcommand{\fiidiii}{F_2^{D(3)}}
\newcommand{\fiidiiiarg}{\fiidiii\,(\beta,\,Q^2,\,x)}
\newcommand{\n}{1.19\pm 0.06 (stat.) \pm0.07 (syst.)}
\newcommand{\nz}{1.30\pm 0.08 (stat.)^{+0.08}_{-0.14} (syst.)}
\newcommand{\fiidiiiful}{F_2^{D(4)}\,(\beta,\,Q^2,\,x,\,t)}
\newcommand{\fiipom}{\tilde F_2^D}
\newcommand{\ALPHA}{1.10\pm0.03 (stat.) \pm0.04 (syst.)}
\newcommand{\ALPHAZ}{1.15\pm0.04 (stat.)^{+0.04}_{-0.07} (syst.)}
\newcommand{\fiipomarg}{\fiipom\,(\beta,\,Q^2)}
\newcommand{\pomflux}{f_{\pom / p}}
\newcommand{\nxpom}{1.19\pm 0.06 (stat.) \pm0.07 (syst.)}
\newcommand {\gapprox}
   {\raisebox{-0.7ex}{$\stackrel {\textstyle>}{\sim}$}}
\newcommand {\lapprox}
   {\raisebox{-0.7ex}{$\stackrel {\textstyle<}{\sim}$}}
\def\gsim{\,\lower.25ex\hbox{$\scriptstyle\sim$}\kern-1.30ex%
\raise 0.55ex\hbox{$\scriptstyle >$}\,}
\def\lsim{\,\lower.25ex\hbox{$\scriptstyle\sim$}\kern-1.30ex%
\raise 0.55ex\hbox{$\scriptstyle <$}\,}
\newcommand{\pomfluxarg}{f_{\pom / p}\,(x_\pom)}
\newcommand{\dsf}{\mbox{$F_2^{D(3)}$}}
\newcommand{\dsfva}{\mbox{$F_2^{D(3)}(\beta,Q^2,x_{I\!\!P})$}}
\newcommand{\dsfvb}{\mbox{$F_2^{D(3)}(\beta,Q^2,x)$}}
\newcommand{\dsfpom}{$F_2^{I\!\!P}$}
\newcommand{\gap}{\stackrel{>}{\sim}}
\newcommand{\lap}{\stackrel{<}{\sim}}
\newcommand{\fem}{$F_2^{em}$}
\newcommand{\tsnmp}{$\tilde{\sigma}_{NC}(e^{\mp})$}
\newcommand{\tsnm}{$\tilde{\sigma}_{NC}(e^-)$}
\newcommand{\tsnp}{$\tilde{\sigma}_{NC}(e^+)$}
\newcommand{\st}{$\star$}
\newcommand{\sst}{$\star \star$}
\newcommand{\ssst}{$\star \star \star$}
\newcommand{\sssst}{$\star \star \star \star$}
\newcommand{\tw}{\theta_W}
\newcommand{\sw}{\sin{\theta_W}}
\newcommand{\cw}{\cos{\theta_W}}
\newcommand{\sww}{\sin^2{\theta_W}}
\newcommand{\cww}{\cos^2{\theta_W}}
\newcommand{\trm}{m_{\perp}}
\newcommand{\trp}{p_{\perp}}
\newcommand{\trmm}{m_{\perp}^2}
\newcommand{\trpp}{p_{\perp}^2}
\newcommand{\alp}{\alpha_s}

\newcommand{\alps}{\alpha_s}
\newcommand{\sqrts}{$\sqrt{s}$}
\newcommand{\LO}{$O(\alpha_s^0)$}
\newcommand{\Oa}{$O(\alpha_s)$}
\newcommand{\Oaa}{$O(\alpha_s^2)$}
\newcommand{\PT}{p_{\perp}}
\newcommand{\JPSI}{J/\psi}
\newcommand{\sh}{\hat{s}}
\newcommand{\uh}{\hat{u}}
\newcommand{\MP}{m_{J/\psi}}
\newcommand{\PO}{I\!\!P}
\newcommand{\xbj}{x}
\newcommand{\xpom}{x_{\PO}}
\newcommand{\ttbs}{\char'134}
\newcommand{\xpomlo}{3\times10^{-4}}  
\newcommand{\xpomup}{0.05}  
\newcommand{\dgr}{^\circ}
\newcommand{\pbarnt}{\,\mbox{{\rm pb$^{-1}$}}}
\newcommand{\gev}{\,\mbox{GeV}}
\newcommand{\mev}{\,\mbox{MeV}}
\newcommand{\GeV}{\,\mbox{GeV}}
\newcommand{\WBoson}{\mbox{$W$}}
\newcommand{\fbarn}{\,\mbox{{\rm fb}}}
\newcommand{\fbarnt}{\,\mbox{{\rm fb$^{-1}$}}}
\newcommand{\dsdx}[1]{$d\sigma\!/\!d #1\,$}
\newcommand{\eV}{\mbox{e\hspace{-0.08em}V}}
%
%
\newcommand{\mc}{MC }
\newcommand{\mcs}{MCs }

\newcommand{\kt}{\ensuremath{k_T}}
\newcommand{\pperp}{\ensuremath{p_\perp}}
\newcommand{\ptg}{\ensuremath{\vec{p}_T^{~\gamma}}}
\newcommand{\ptjet}{\ensuremath{\vec{p}_T^{~jet}}}
\newcommand{\Dphi}{\ensuremath{\varDelta\varPhi}}
\newcommand{\Etg}{\ensuremath{E_T^\gamma}}
\newcommand{\Eg}{\ensuremath{E^\gamma}}
\newcommand{\etg}{\ensuremath{E_T^\gamma}}
\newcommand{\etgjet}{\ensuremath{E_T^{~\gamma\mathrm{-jet}}}}
\newcommand{\etag}{\ensuremath{\eta^\gamma}}
\newcommand{\etj}{\ensuremath{E_T^{\rm{jet}}}}
\newcommand{\Ptj}{\ensuremath{p_T^{\rm{jet}}}}
\newcommand{\ptj}{\ensuremath{p_T^{\rm{jet}}}}
\newcommand{\Etj}{\ensuremath{E_T^{\rm{{jet}}}}}
\newcommand{\etaj}{\ensuremath{\eta^{\rm{{jet}}}}}
\newcommand{\etjet}{\ensuremath{E_T^{\rm{{jet}}}}}
\newcommand{\etajet}{\ensuremath{\eta^{\rm{{jet}}}}}
\newcommand{\xg}{\ensuremath{x_\gamma}}
\newcommand{\xp}{\ensuremath{x_{p}}}
\newcommand{\xglo}{\ensuremath{x_\gamma^{LO}}}
\newcommand{\xplo}{\ensuremath{x_{p}^{LO}}}

\newcommand{\qsq}{\ensuremath{Q^2} }
\newcommand{\gevsq}{\ensuremath{\mathrm{GeV}^2} }
\newcommand{\et}{\ensuremath{E_t^*} }
\newcommand{\rap}{\ensuremath{\eta^*} }
\newcommand{\gp}{\ensuremath{\gamma^*}p }
\newcommand{\dsiget}{\ensuremath{{\rm d}\sigma_{ep}/{\rm d}E_t^*} }
\newcommand{\dsigrap}{\ensuremath{{\rm d}\sigma_{ep}/{\rm d}\eta^*} }

\newcommand{\dstar}{\ensuremath{D^*}}
\newcommand{\dstarp}{\ensuremath{D^{*+}}}
\newcommand{\dstarm}{\ensuremath{D^{*-}}}
\newcommand{\dstarpm}{\ensuremath{D^{*\pm}}}
\newcommand{\zDs}{\ensuremath{z(\dstar )}}
\newcommand{\Wgp}{\ensuremath{W_{\gamma p}}}
\newcommand{\ptds}{\ensuremath{p_t(\dstar )}}
\newcommand{\etads}{\ensuremath{\eta(\dstar )}}
\newcommand{\ptjn}[1]{\ensuremath{p_t(\mbox{jet$_{#1}$})}}

\newcommand{\detadsj}{\ensuremath{\eta(\dstar )\, \mbox{-}\, \etaj}}
\newcommand{\matrixA}{\ensuremath{\textbf{A}}}

\def\Journal#1#2#3#4{{#1} {\bf #2} (#3) #4}
\def\NCA{\em Nuovo Cimento}
\def\NIM{\em Nucl. Instrum. Methods}
\def\NIMA{{\em Nucl. Instrum. Methods} {\bf A}}
\def\NPB{{\em Nucl. Phys.}   {\bf B}}
\def\PLB{{\em Phys. Lett.}   {\bf B}}
\def\PRL{\em Phys. Rev. Lett.}
\def\PRD{{\em Phys. Rev.}    {\bf D}}
\def\ZPC{{\em Z. Phys.}      {\bf C}}
\def\EJC{{\em Eur. Phys. J.} {\bf C}}
\def\CPC{\em Comp. Phys. Commun.}

\begin{titlepage}

\noindent
\begin{flushleft}
{\tt DESY 09-135    \hfill    ISSN 0418-9833} \\
{\tt January 2010}                  \\
\end{flushleft}

\noindent
\noindent

\vspace{2cm}
\begin{center}
\begin{Large}

{\bf  Prompt Photons  in Photoproduction at HERA }

\vspace{2cm}

H1 Collaboration

\end{Large}
\end{center}

\vspace{2cm}

\begin{abstract}
\noindent
The production of prompt photons is measured in the  photoproduction regime 
of electron-proton scattering at HERA.
The analysis is based on a  data sample corresponding to a total integrated
 luminosity  of  $340 $~{\rm pb}$^{-1}$ collected by the H1 experiment.
Cross sections are measured for photons with transverse momentum and  pseudorapidity
in the range $6 < \etg < 15 $\GeV{} and $-1.0 < \etag < 2.4$, respectively. 
Cross sections for events with an additional jet are 
measured as a function of the transverse energy and pseudorapidity 
of the jet, and as a function of the fractional momenta  $\xg$ and $\xp$ carried by the partons 
entering the hard scattering process.
The correlation between the photon and the jet is also studied. 
The results are compared with QCD  predictions based on the collinear 
and on the $k_T$ factorisation approaches.

\end{abstract}

\vspace{1.5cm}

\begin{center}
Accepted by \EJC\
\end{center}

\end{titlepage}

%
%
%
\begin{flushleft}


F.D.~Aaron$^{5,49}$,           
M.~Aldaya~Martin$^{11}$,       
C.~Alexa$^{5}$,                
K.~Alimujiang$^{11}$,          
V.~Andreev$^{25}$,             
B.~Antunovic$^{11}$,           
S.~Backovic$^{30}$,            
A.~Baghdasaryan$^{38}$,        
E.~Barrelet$^{29}$,            
W.~Bartel$^{11}$
% ...

\bigskip{\it
 $ ^{1}$ I. Physikalisches Institut der RWTH, Aachen, Germany \\
% ...
$ ^{5}$ National Institute for Physics and Nuclear Engineering (NIPNE) ,
          Bucharest, Romania \\
% ...
$ ^{11}$ DESY, Hamburg, Germany \\
% ...
$ ^{25}$ Lebedev Physical Institute, Moscow, Russia$^{ e}$ \\
% ...
$ ^{29}$ LPNHE, Universit\'{e}s Paris VI and VII, IN2P3-CNRS,
          Paris, France \\
 $ ^{30}$ Faculty of Science, University of Montenegro,
          Podgorica, Montenegro$^{ e}$ \\
% ...
$ ^{38}$ Yerevan Physics Institute, Yerevan, Armenia \\
% ...
$ ^{49}$ Also at Faculty of Physics, University of Bucharest,
          Bucharest, Romania \\
% ...
$ ^e$ Supported by the Deutsche Forschungsgemeinschaft \\
% ...
}
\end{flushleft}
%

\newpage

\section{Introduction}

Isolated photons emerging from the hard subprocess $ep \rightarrow e\gamma X$,
 so called prompt photons, are a powerful probe of the underlying dynamics,
complementary to jets.
 Production of isolated photons with high transverse momentum 
can be calculated in  perturbation theory.
High energy electron-proton scattering is dominated by so-called 
photoproduction
processes, in which a beam lepton emits a quasi-real photon which either 
interacts directly with the proton (direct process) or fluctuates into 
partons which then participate in the hard scattering process (resolved process).
In prompt photon production, the direct process is sensitive to the quark content of the proton 
through the Compton scattering of the exchanged photon with a quark
 ($\gamma q\rightarrow \gamma q$) as depicted in figure~\ref{fig:feynman}a). The resolved process ($q g  \rightarrow \gamma q$) is  sensitive to 
 the partonic structure of both the photon
                      and the proton. 
A typical diagram is shown in figure~\ref{fig:feynman}b).
Figure~\ref{fig:feynman}c) and~\ref{fig:feynman}d) show typical higher order diagrams.

\begin{figure}[Hhh]
  \begin{center}
    \includegraphics[height=0.19\textwidth]{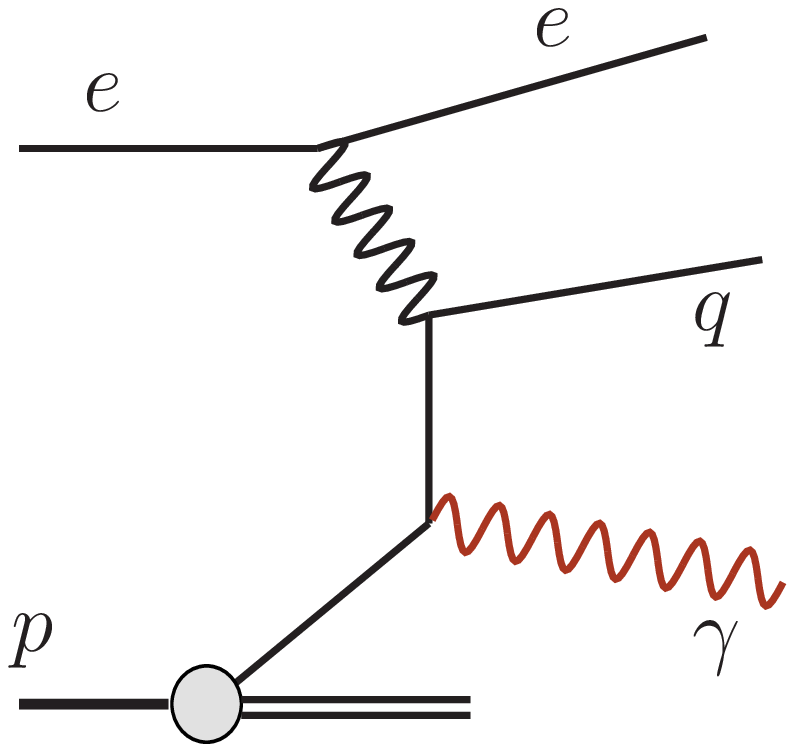}
    \hspace{0.01\textwidth}
    \includegraphics[height=0.19\textwidth]{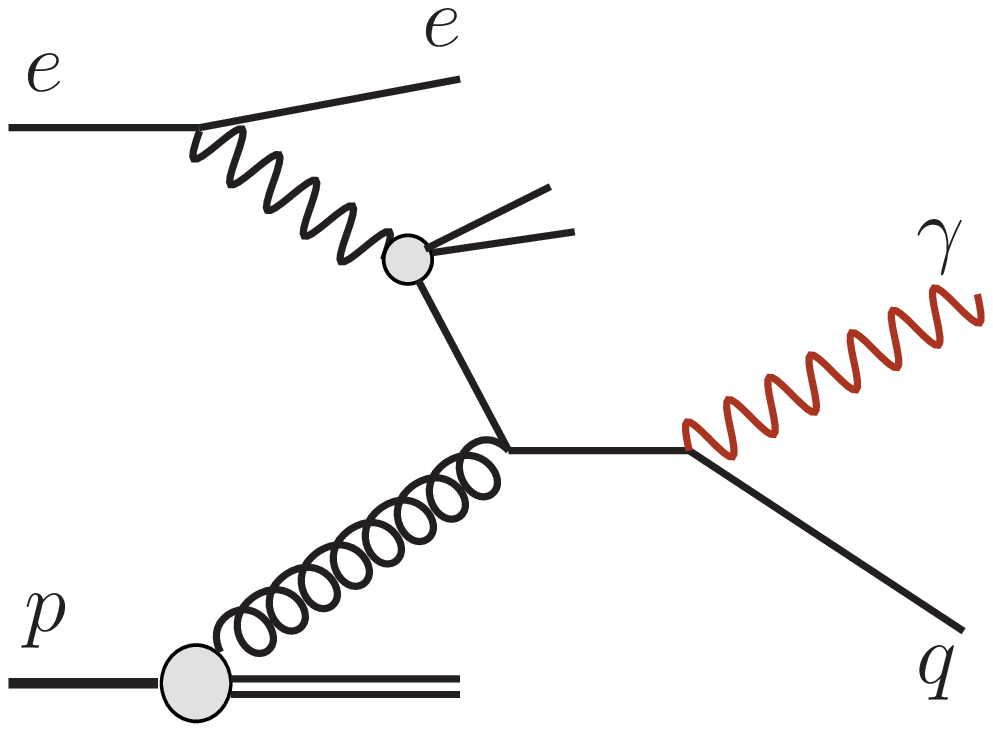}
   \hspace{0.01\textwidth}
    \includegraphics[height=0.19\textwidth]{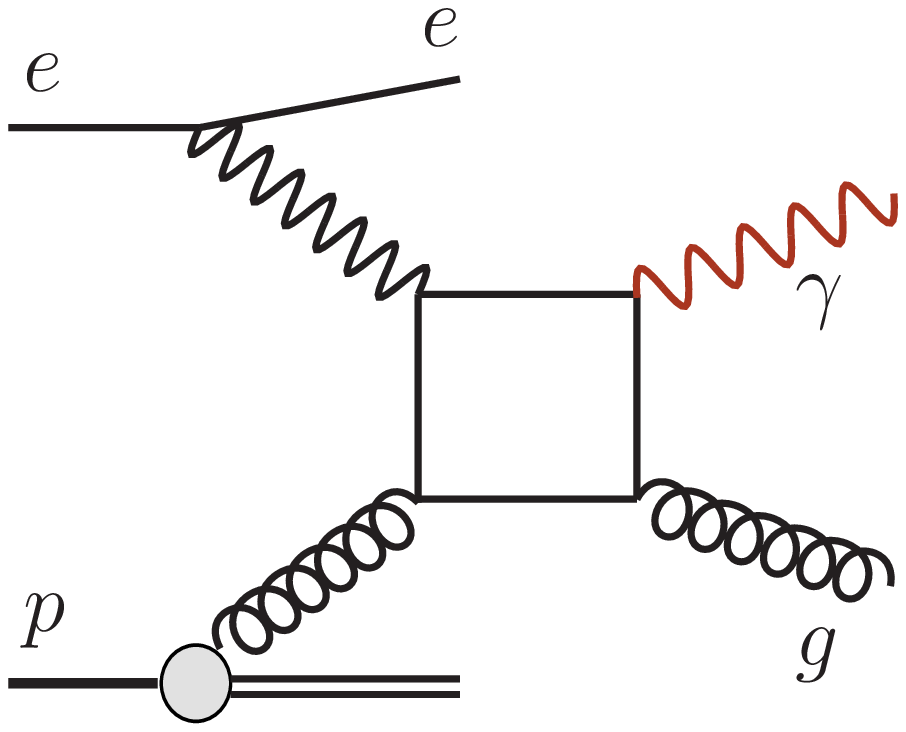}
    \hspace{0.01\textwidth}
    \includegraphics[height=0.19\textwidth]{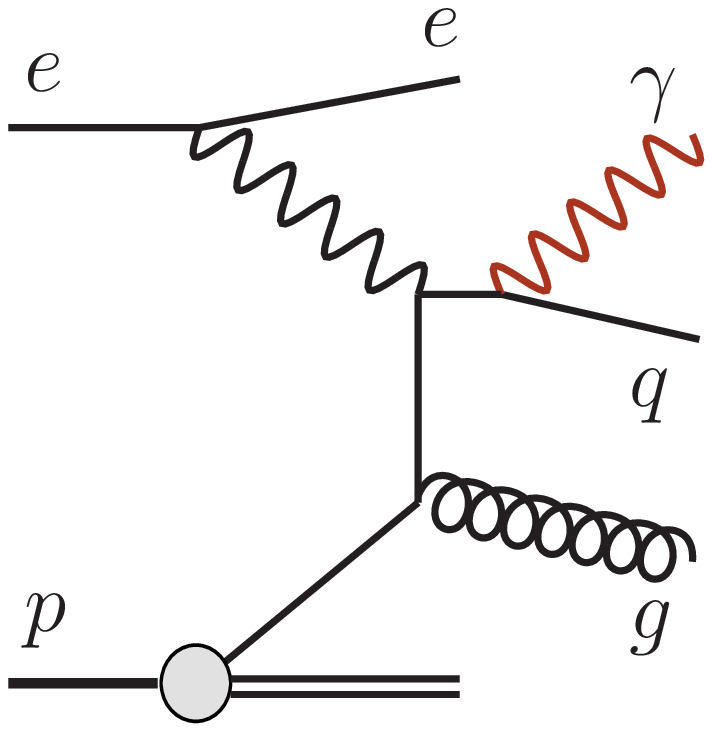}
  \end{center}
  \begin{picture} (0.,0.)
     \setlength{\unitlength}{1.0cm}
     \put ( 0.3, 2.2){a)}
     \put ( 4. ,2.2){b)}
     \put ( 8.6, 2.2){c)}
     \put (12.7, 2.2){d)}
  \end{picture}

    \caption{Examples of diagrams for the production of prompt photons in photoproduction: a) direct and b) resolved interaction with a parton from the proton, c) box diagram and d) radiation of a photon from an outgoing quark.
    }
    \label{fig:feynman}
\end{figure}

The H1 collaboration has previously measured prompt photon 
cross sections in  photoproduction~\cite{H1gamma} and 
 in deep inelastic scattering (DIS)~\cite{H1dis}. 
The ZEUS collaboration has also reported  measurements of  prompt photon production~\cite{ZEUS,ZEUS1,ZEUS2}. 
Both experiments found that in photoproduction the inclusive prompt photon
 cross section
is underestimated by next-to-leading order (NLO) QCD calculations~\cite{heinrich,heinrich2,nlo1}, while there is reasonable agreement for events with a
prompt photon and a jet (photon plus jet).
In DIS, a leading order  QCD calculation~\cite{thomas} significantly underestimates the production of 
isolated photons  and  of photons plus jets. NLO predictions~\cite{kramer} are only available for the latter and also underestimate the cross section.

This paper presents results of a measurement of  prompt  photons in photoproduction. 
The data used for the measurement were collected with the H1 detector 
in the period from $2004$ to $2007$ and correspond to a total integrated luminosity of  $340$~{\rm pb}$^{-1}$.
This
amounts to an increase in statistics by a factor of three compared to the previous 
measurement~\cite{H1gamma}.
During this data taking period HERA collided positrons or electrons\footnote
{Unless  otherwise stated,  the term electron  refers to both the electron and the positron.} 
of energy $E_e = 27.6~{\rm GeV}$ with protons of energy $E_p = 920~{\rm GeV}$ corresponding to a  centre-of-mass energy of $\sqrt{s} = 319~{\rm GeV}$. 

Isolated photons with  transverse energy $6< E_T^\gamma < 15$~GeV
and  pseudorapidity\footnote{
The  pseudorapidity is related to the polar angle $\theta$ as $\eta =
- \ln \tan(\theta/2)$, where $\theta$ is measured with respect to the
direction of the outgoing proton beam (forward direction).
}
   $-1.0 < \eta^\gamma < 2.4$
are measured  in events with the inelasticity $y$ in the range $0.1<y< 0.7$.
This extends the phase space of previous measurements at HERA
 towards larger pseudorapidities of the photon and to
smaller event inelasticities. 

The main background is due to photons produced in hadron decays.
For its discrimination  from prompt photons, 
 various shower shape variables are  used. 
Differential cross sections are presented as a function of the 
transverse energy and pseudorapidity of the photon.
For the photon plus jet sample, differential cross sections are measured 
 as a function of 
transverse energy and pseudorapidity of the photon and the jet and the momentum 
fractions $\xg$ and $\xp$ carried by the participating parton in the photon and the proton, 
respectively. Azimuthal angle and transverse momentum correlations between the 
photon and the jet are also studied.
The cross sections are compared to QCD calculations based on collinear 
factorisation in NLO~\cite{heinrich,heinrich2} and to calculations based on the  
$k_T$ factorisation approach~\cite{zotov}.

\section{ Theoretical Predictions}
\label{sect:theo}

The calculation by Fontannaz, Guillet and Heinrich (FGH)~\cite{heinrich,heinrich2} based on the collinear factorisation approach includes the leading order direct and resolved 
processes  $\gamma q \rightarrow \gamma q$ and their NLO corrections. 
Besides the production of a prompt photon in the hard interaction, photons may originate from the fragmentation of a high momentum quark or gluon in the final state.
The fragmentation process, described by a fragmentation function,  is included  in the calculation as 
well as the
 direct box diagram as shown in figure~\ref{fig:feynman}c).
The contribution from quark to photon fragmentation to the total cross section of isolated photons is at the level of $4\%$. The contribution from the box diagram amounts to about $10\%$ on average.
The calculation uses the  parton density functions (PDFs)
CTEQ6L~\cite{cteq6l} for the proton and AFG04\cite{afg} for  the photon.
The scales for renormalisation $\mu_R$ and factorisation $\mu_F$,
 are chosen to be
$\mu_R = \mu_F  =E_T^\gamma$. 
The NLO corrections to the LO cross section are significant for the inclusive sample. 
They increase the predicted cross section by a 
factor $1.15-1.42$,
the corrections being largest  at low $\etg$ and large $\etag$.
For the photon plus jet sample  the corrections are much smaller and below $10\%$ on average.

The leading order predictions of Lipatov and Zotov (LZ)~\cite{zotov}  are based on the $k_T$ factorisation approach.
The calculation uses the unintegrated quark and gluon densities of the photon and the proton using the Kimber-Martin-Ryskin (KMR) 
prescription~\cite{kimber} with  the GRV para\-meterisations for the collinear quark and gluon densities~\cite{grv,grv2}.
The $k_T$ factorisation approach is expected to account for the main part of the collinear higher order QCD corrections~\cite{zotov}.
Direct and resolved processes are considered in the calculation, but contributions from fragmentation  and from the box diagram 
are neglected.

To ensure isolation of the photon,  the total  transverse energy within a 
cone of radius one in  the
pseudorapidity - azimuthal angle plane 
surrounding the prompt photon, excluding its own energy, is 
required to be below $10\%$ of $\etg$ in both calculations.
This requirement  slightly differs from the one used in the data  analysis as described in section~\ref{sect:sel}. 

The theoretical predictions are compared to the data after a correction for multi parton interactions, for hadronisation  effects  and for the different definition of the isolation of the photon.
The total correction factors $f_{corr}$ are determined with the signal MC described below 
as the ratios of the cross sections on hadron level with multi parton interactions and the 
data isolation criteria, to the cross sections on parton level without multi parton interactions and using 
the cone cut for the isolation of the photon.
The correction factors are calculated for each bin using the event generators PYTHIA~\cite{PYTHIA} and HERWIG\cite{herwig} which have a different
model for hadronisation. 
The arithmetic means of the two correction factors  are used, while half of the difference between the two models is taken as the error. 
The correction factors for the total inclusive cross section range from $0.84$ to $0.99$  with an average of $0.9$.
They are largest for low $\Etg$ and in the forward direction, where the photon isolation 
is most sensitive to hadronisation and to multi parton interactions. 
The uncertainty of the corrections is typically $8\%$.

The leading order MC generator PYTHIA~$6.2$~\cite{PYTHIA}  is used in this analysis for the 
prediction of the signal.
The simulation of multi parton interactions~\cite{mia1,mia2} is included.
  The  hard partonic interaction is calculated in LO QCD and
higher order QCD radiation is modelled using initial and final state parton showers in the leading log approximation~\cite{Bengtsson:1987rw}. 
The fragmentation into hadrons is simulated in PYTHIA by the Lund string model~\cite{lund}.
The simulated signal contains contributions from direct (figure~\ref{fig:feynman}a) and 
resolved (figure~\ref{fig:feynman}b)  production of prompt photons including QED radiation.
In addition, processes with two hard partons in the final state (figure~\ref{fig:feynman}d) are simulated.  
The simulations use the  parton densities CTEQ6L~\cite{cteq6l} for the  proton and SASG-1D~\cite{sas} for the photon.
Different parton density functions  for the proton (CTEQ5L~\cite{cteq5l} and MRST04~\cite{mrst04}) and the photon (GRV~\cite{grv} and AFG04) are used
to estimate the influence of the parton densities on the predicted 
cross section, which  varies by at most $10\%$, mainly due to changes of the proton PDF.
The multi parton interactions  reduce the total inclusive cross section  by 
$6\%$ on average.
The uncertainty of the correction for multi parton interactions is estimated by changing the default 
parameter for the effective minimum transverse momentum for multi parton interactions in PYTHIA (PARP(81))
 from $1.9$~GeV to $1.6$~GeV and $2.2$~GeV, respectively.

To estimate the uncertainty of the hadronisation correction, 
the HERWIG~\cite{herwig} generator is also used to model the  
prompt photon signal.
HERWIG simulates the fragmentation into hadrons  through the decay of 
colourless parton clusters.

Background to the analysis of prompt photons mainly arises from energetic photons from the  decay  of hadrons like $\pi^0$ and $\eta$ in photoproduction events, which constitute more than $90\%$ of the total background prediction. 
Direct and resolved photoproduction  of di-jet events used to study the background is simulated with PYTHIA.

All generated events are passed through a GEANT~\cite{geant} based simulation of the H1 detector which takes into account the different data taking periods, and are subject to the same reconstruction and analysis chain as the data.

\section{ H1 Detector}

A detailed description of the H1 detector can be found in \cite{H1det}. 
In the following, only  detector components relevant to this analysis are 
briefly discussed. The origin of the H1 coordinate system is the nominal $ep$ 
interaction point, with the direction of the proton
beam defining the positive $z$-axis (forward direction). 
Transverse momenta are measured in the $x$-$y$~plane. Polar ($\theta$) and azimuthal ($\phi$) angles are measured with respect to this reference system.
\par
In the central region (\mbox{$15^\circ\!<\!\theta\!<\!165^\circ$}) the interaction point is surrounded by the central tracking
system (CTD)   , which consists of a silicon vertex detector~\cite{pitzl} and drift chambers
all operated within a solenoidal magnetic
field of $1.16 \rm\ T$.
The forward tracking detector  and the backward proportional chamber 
measure tracks of charged particles at smaller \mbox{($7^\circ\!<\!\theta\!<\!25^\circ$)} and larger \mbox{($155^\circ\!<\!\theta\!<\!175^\circ$)} polar angles
than the central tracker, respectively.
In each event the $ep$ interaction vertex is reconstructed from the charged tracks.
In the polar angular region ($11^\circ < \theta < 169^\circ$) an additional 
track signature is obtained from a set of five cylindrical
 multi-wire proportional chambers (CIP2k)~\cite{cip2000}.

\par 
The liquid argon (LAr) sampling calorimeter~\cite{Andrieu:1993kh} surrounds the tracking
chambers. It has a  polar angle coverage of 
\mbox{$4^\circ\!<\!\theta\!<\!154^\circ$} and full azimuthal acceptance.
 It consists of an inner electromagnetic section with lead absorbers and an outer hadronic section with steel absorbers.
The calorimeter is divided into eight wheels  along the beam axis.
The electromagnetic and the hadronic sections are highly segmented in the transverse and the longitudinal directions.
Electromagnetic shower energies are measured with a precision of
$\sigma(E)/E=12\%/\sqrt{E/\gev}\oplus 1\%$ and hadronic energies with
$\sigma(E)/E=50\%/\sqrt{E/\gev}\oplus 2\%$, as determined in test beam experiments~\cite{Andrieu:1994yn,h1testbeam}. 
In the backward
region (\mbox{$153^\circ\!<\!\theta\!<\!178^\circ$}), particle energies are measured
by a lead-scintillating fibre spaghetti calorimeter (SpaCal)~\cite{Appuhn:1996na}. 
\par 
The luminosity is determined from the rate of the  Bethe-Heitler process 
$ep\!\rightarrow\!ep\gamma$, measured using a photon detector located close 
to the beam pipe at $z=-103$~m.
\par
The LAr calorimeter provides the trigger~\cite{Trig} for the events in this analysis.
The hardware trigger is complemented by a software trigger requiring an electromagnetic cluster in the LAr calorimeter with a transverse energy $\Etg>5$~GeV.  
The combined trigger efficiency is about $85$\% at $\Etg$ of $6$~GeV rising to  above
$95$\% for $\Etg >$ $7$~GeV. 

\section{ Experimental Method}
\subsection{ Event Selection and Reconstruction}
\label{sect:sel}

Events are selected with a  photon candidate  in the LAr calorimeter
 of transverse energy  $6 < \etg < 15 $~GeV and pseudorapidity $-1.0 < \etag < 2.4$. 
Photon candidates are defined as  compact clusters in the electromagnetic section of the LAr 
calorimeter with no matching
signals in the CIP2k.
The CIP2k veto rejects candidates, if there is a signal  in at least two 
layers of the CIP2k close to the expected hit position.
In addition, a track veto is applied for $\theta>45^{\circ}$. It rejects candidates,
 if a  track in the CTD  extrapolated to the LAr calorimeter front face matches
the electromagnetic cluster with a distance of closest approach to the cluster's barycentre of less than $15$~cm. 
Photon candidates are also rejected if they are  close to  
inactive regions between calorimeter modules.

Neutral current (NC) deep-inelastic scattering (DIS) events are suppressed by 
rejecting events with an electron candidate not previously identified as photon candidate.
Electron candidates are defined as compact electromagnetic clusters in the SpaCal or in the LAr calorimeter.
In the LAr calorimeter the candidates are required to have an associated track with a distance of closest approach of less than $12$~cm. 
The electron suppression restricts the sample to NC events where the
scattered electron escapes along the beam pipe in the negative $z$
direction. The low electron scattering angle of such events
corresponds to 
virtualities of the exchanged photon in the range $Q^2<4$~GeV$^2$.
In photoproduction the inelasticity $y$ is expressed as $y = W^2/s$, where $W$ is the $\gamma p$ centre of mass energy.
In this analysis $y$ is evaluated as $y_{h}=\Sigma(E-p_z)/2E_e$,
where the sum runs over all measured  final state particles with energy $E$ and longitudinal momentum $p_z$. 
The inelasticity is restricted to $0.1<y_h<0.7$.
The cut at low $y_h$ removes residual beam gas background and the higher cut on $y_h$ removes background 
from DIS events including events with prompt photons and events where the scattered electron is misidentified as a photon. 
This background is below $1.5$\% in the final sample and is considered as a systematic uncertainty.

In order to remove background events from non-$ep$ sources, 
at least two tracks are required in the central tracker, assuring a good reconstruction of the
longitudinal event vertex position which is required to be within  $40$~cm around the nominal 
interaction point.
In addition, topological filters and timing vetoes are applied to remove cosmic muons and beam induced background. 

The shape of the photon cluster candidate is used to further reduce the background.
The transverse\footnote{In the context of the cluster shape analysis the transverse plane is defined as perpendicular
  to the direction of the photon candidate.} radius  $R_T$ of the
  photon candidate
   is defined as the square root of the second central transverse moment  
$R_T=\sqrt{\mu_2}$, where the $k$'th central transverse moment of the calorimeter cells distribution is given by
$\mu_k= \langle |\vec{r}- \langle\vec{r}\rangle |^k \rangle $. 
Here, $\vec{r}$ is the transverse projection of a cell 
position and the averages are calculated taking into account the cell
energies as weight factors.
The requirement $R_T < 6$~cm reduces background from 
neutral hadrons that decay into multiple photons.
In most cases such decay photons  are merged into one electromagnetic cluster, which tends to have a wider
 transverse  spread than that of a single photon. 

For events where a second electromagnetic cluster is found,
the invariant mass $M_{\gamma\gamma}$ of the photon candidate cluster,  combined with the closest neighbouring 
electromagnetic   cluster with an energy above $80$~\mev, is reconstructed.
Photon candidates from $\pi^0$ decays where the two decay photons are reconstructed in separate clusters
are rejected requiring $M_{\gamma\gamma} > 300 \mev$.

\par
Tracks and calorimeter energy deposits not previously identified as photon candidate 
are used to form combined cluster-track objects.
The photon candidate and  the 
cluster-track objects are combined into massless jets using 
the inclusive 
 $k_T$ algorithm~\cite{jetalgo} with the separation parameter $R_0$  set to $1$.
 Jets are reconstructed  in the pseudorapidity range $-2.0<\etaj<3.0$  with a  transverse momentum of
$\Etj > 4$~GeV.
Due to the harder kinematical cuts for the photon candidate there is always a jet containing the photon candidate called the photon-jet. 
 All other jets are classified as hadronic jets.
To ensure isolation of the photon, the  
fraction $z=\etg/\etgjet$ of the transverse energy of the photon-jet carried by the photon candidate has to  be larger than $0.9$. Here, $\etgjet$ is the transverse energy of the photon-jet.
This isolation requirement largely suppresses background from
photons produced in the hadron decay cascade.
Only events with exactly one isolated photon candidate are accepted.

For the photon plus jet sample, events are selected with a photon candidate and  at least one  hadronic jet 
with  $-1.3<\etaj<2.3$.
If more than one hadronic jet is selected, the one with the highest $\Etj$ is used.

Four additional observables are defined for the photon plus jet sample which are sensitive to the underlying partonic process:
\begin{itemize}
\item The estimators $\xglo$ and $\xplo$, which in the LO approximation correspond to the 
 longitudinal momentum fractions of the partons in the photon and the proton, respectively,
are defined as
\beq
\xglo = \frac{\etg(e^{-\etaj} + e^{-\etag})}{(2y_hE_e)}~~~~\mbox{and}~~~~\xplo = \frac{\etg(e^{\etaj} + e^{\etag})}{(2E_p)}~\mbox{.}
\eeq 
These definitions~\cite{heinrich3,afg94}  reduce infrared sensitivity for $x_\gamma\rightarrow 1$ 
compared to the conventional definition of $x_\gamma = ( \etg e^{-\etag} +  \etj e^{-\etaj})/(2yE_e)$. 
The above definitions make  use of the energy of the photon only, which has  a better resolution than the  energy of the jet.
However, $\xglo$ and $\xplo$ may become larger than unity. 
\item Two observables $\pperp$ and $\Dphi$ describe the transverse 
correlation between the photon and the jet,
$\Dphi$ is the azimuthal difference between the photon and the jet, and 
$p_\perp$ is the photon momentum component 
perpendicular to the jet direction in the transverse plane
\beq
\pperp \equiv\frac{ \mid\ptg \times \ptjet \mid}{\mid \ptjet\mid} = \etg \cdot \sin{\Dphi}.
\eeq
At leading order the prompt photon and the jet are back-to-back and $\pperp$ equals zero for direct processes. 
The observable $\Dphi$ is strongly correlated with $\pperp$, but is less sensitive to the energies of the photon and the jet.
\end{itemize}

The $y_h$, $\etg$,  $z$ and $\etajet$  distributions of events with an isolated photon candidate 
are shown in figure~\ref{fig:control} together with the
 MC predictions from PYTHIA for the signal and the background. 
The signal (background) prediction is scaled by a factor $1.45$ ($1.7$) on average.
  The scaling factors vary  as a function of $\eta$ as  suggested by the cross section measurement (section~\ref{sect:res}).
In all distributions the data are described within errors by the scaled MC predictions.
At this stage of the analysis there is still a significant contribution of background from the
decay products of neutral mesons. 

\subsection{ Photon Signal Extraction}
\label{sect:phot}

The photon signal is extracted from the sample with photon candidates by means of a shower shape analysis based on the method described in  \cite{H1dis,carsten}.
It uses the following six shower shape variables calculated from the measurements of the individual cells composing the 
cluster:

\begin{itemize}
\item The transverse radius of the cluster, $R_T$.
\item The transverse symmetry,  which is the ratio of 
the spread  of the transverse
 cell distributions along the two principal axes.
Single photon clusters are  expected to be more 
  symmetric than  multi-photon clusters.
\item The transverse kurtosis,  defined as
$K_T=\mu_4/(\mu_2)^2 - 3$, with $\mu_2$ and $\mu_4$ the second and the fourth moment of the transverse energy distribution.
\item The first layer fraction, defined as the fraction of the cluster's  energy detected in the first calorimeter layer. 
\item The hot core fraction, being the fraction of the energy of the electromagnetic cluster contained in the hot core of the cluster.
It is defined as the energy fraction in four to twelve
 contiguous cells in the first two calorimeter layers, depending on the polar angle.
 The cells include the most energetic cell and are 
chosen to maximise the energy.  
\item The hottest cell fraction,  which is the fraction of the energy of the electromagnetic cluster contained in the cell with the largest energy deposit.
\end{itemize}

The distributions of the shower shape variables are shown  in figure~\ref{fig:showershapes} for the prompt
photon candidates with the kinematic cuts as defined above.
The shaded band shows the systematic uncertainty assigned to the description of the shower shapes as described in section~\ref{sect:syst}.
The data are compared with the sum of the background and the signal  MC 
distributions, which describe the data within the systematic error.

In order to discriminate between  signal and background, 
probability density functions   
$p^{i}_{\gamma}$ for the signal and  $p^{i}_{bg}$ for the background are 
defined for each of the six shower shape variables $i$.
Simulated events for the signal and the background are used to determine $p^i_{\gamma}$ and  $p^{i}_{bg}$.
The photon and background probability densities are
 taken as the product of the respective shower shape densities with the method described in \cite{likeli}. 
For each  event a discriminator $D$ is formed.
It is defined as the photon probability density divided by the sum of
 the probability densities for photons and background.
 Figure~\ref{fig:discriminant}
 shows  an example of the discriminator distribution for the range $0.94<\etag<1.42$ and 
four different bins in $\etg$.  
The discriminator has in general larger values for prompt photons than for
the decay photons. The separation power is decreasing with increasing $\etg$.  
The sum of the MC predictions describes the data within 
the systematic uncertainty of the shower shapes.

Additional event samples 
are used for the determination of systematic errors related to the cluster shapes.
The first   sample, containing Bethe Heitler events,  $ep\rightarrow e\gamma p$, consists of events with an electron reconstructed 
in the LAr 
calorimeter, a photon in the SpaCal and nothing else in the detector. 
The second, complementary  sample, in addition containing
deeply-virtual Compton scattering \cite{DVCS} events, is 
selected by requiring  an electron in the SpaCal, a photon in the LAr calorimeter  and no other particle in the detector. 
These independent event selections, denoted BH and DVCS respectively,
provide a clean sample of electromagnetic clusters at low transverse energies in the LAr 
calorimeter and are used to study the description of the shower shapes of the photons. 
A third sample is used to monitor the description of the shower shapes of clusters initiated by the decay of neutral hadrons. 
This sample, denoted BG, is background enhanced by selecting events with  the inverted isolation criteria $z=\etg/\etgjet<0.9$ and no cut on the transverse radius of the photon candidate.

\subsection{ Cross Section Determination}

A  regularised unfolding
procedure~\cite{phillips,thikanov,blobel,tunfold,krzys} is used to
relate distributions $\vec{y}_{\mathrm{rec}}$  of 
reconstructed variables (input bins) to distributions
$\vec{x}_{\mathrm{true}}$ of variables on hadron level (output bins), to
 determine the fractions of signal 
and background and to correct the data for the detector acceptance.
 The unfolding  matrix \matrixA{}  relates the two vectors,  
$\matrixA \vec{x}_{\mathrm{true}} =\vec{y}_{\mathrm{rec}}$.
 Further details on the method can be found in in~\cite{krzys}
and are summarised in appendix~\ref{sect:unfold}.

The input is binned in three dimensions in the reconstructed quantities $\Etg$, $\etag$ and $D$; the latter
 allows the discrimination of signal and background.
The output of the unfolding procedure contains the number of signal events $N_{sig}$ in  $\Etg$-$\etag$ bins  on hadron level and
the amount of background events in any of the input bins. 
Additional underflow and overflow bins are defined for each output  variable. 
Therefore the unfolding matrix \matrixA{} also includes migrations into or out of the  phase space of the measurement. 
It is computed using signal and background PYTHIA simulation.
For measurements including jet-related variables, both the  input and the output is additionally binned
in some variable $u$, where $u$ is $\etjet$, $\etajet$, $\xg$, $\xp$, $\Dphi$ or $\pperp$.

The stability of the unfolding procedure is checked  by varying the number of input bins and changing the bin boundaries.
The results from the unfolding procedure are compared to a bin-by-bin  correction method. Agreement is seen within errors for 
most of the analysis bins.

Cross sections are presented for $Q^2<1$~GeV$^2$.
The extracted number of signal events in each bin $N_{sig} $ is corrected for 
a small contribution of DIS events at virtualities $1<Q^2<4$~GeV$^2$. For this kinematic region the scattered electron has a non-negligible probability to escape detection. If such events contain in addition photons at high transverse momentum, their signatures are very similar to the signal process.
The corresponding correction factor $f_{DIS}$ is determined with the PYTHIA signal MC and is found to be above $0.98$ for most of the analysis bins.
The bin-averaged double differential cross section on hadron level is obtained as
\beq
\frac{\mbox{d}^2\sigma}{\mbox{d}\Etg\mbox{d}\etag} = \frac{N_{sig}\cdot f_{DIS}}{(\mathcal{L}\cdot \Delta \Etg \Delta \etag)},
\eeq
where $\mathcal{L}$ is the luminosity, $\Delta \Etg$ ($\Delta \etag$) is the bin width in $\Etg$ ($\etag$) and $N_{sig}$ corresponds to the number of signal events in the bin $\Etg$-$\etag$.
Single differential cross sections as a function of $\Etg$ ($\etag$) 
are then obtained by summing bins of the double differential cross sections in $\etag$ ($\Etg$), taking into account the respective bin widths.
The total inclusive cross section is obtained by summing the 
measured double differential cross section over all analysis bins.
The differential cross sections in bins of some jet-related variable $u$ is obtained by
unfolding triple-differential cross sections in $\Etg$, $\etag$ and $u$, which then are summed over the bins in $\etg$ and $\etag$.
For the calculation of cross section uncertainties, correlations between bins are taken into account.

\subsection{ Systematic Uncertainties}
\label{sect:syst}

The following experimental uncertainties are considered:
\begin{itemize}
\item The measured shower shape variables in 
the  DVCS and BH event samples defined in section~\ref{sect:phot} are compared to MC simulations. 
The uncertainty on the shower shape simulation for the photon is estimated by varying the discriminating 
variables within the limits deduced from the differences  between data and simulation.
The uncertainty of the description of the background composition  and the 
shower shapes of neutral hadrons is obtained accordingly by comparing the shower shapes of the BG event sample with the background MC from PYTHIA.
The resulting variation of the total inclusive cross section is $11$\%. 
The uncertainty varies between $10$\% and $25$\% for
 the single differential cross sections increasing towards large $\etag$.
\item  An uncertainty of  $1$\% is attributed to the energy of the  photon\cite{DVCS}
  and an uncertainty of $3$~mrad  to the 
measurement of the scattering angle~\cite{H1dis} for events with   $\etag <1.4$.
For $\etag >1.4$ the uncertainty  is $4$\% on the energy scale and  $4$~mrad  on the scattering angle.
The resulting error on the total inclusive cross section is $\pm1.5$\%.

\item A $2$\% uncertainty is attributed to the measurement of the hadronic energy~\cite{krzys}.
 The corresponding uncertainty of the total cross section $\pm1$\%.

\item An uncertainty of $\pm3$\% is attributed to the determination of the trigger efficiency.

\item 
The uncertainty on the CIP2k and track veto efficiency results in an error of $\pm 2.5$\% on the total inclusive cross section.

\item Background from DIS events leads to a systematic uncertainty of $1.5\%$.

\item  An uncertainty in the description of the dead material in the simulation is accounted for by
varying the probability of photon conversion before the calorimeter by  $\pm 10$\%. For polar angles 
$ \theta < 20^\circ$ it is varied by  $\pm 30$\% because of more dead material in the forward region. This results in a $1$\% uncertainty for the cross section measurements in the central region and $3$\% in the 
most forward $\etag$ bin.

\item The ratio of resolved to direct photoproduction events in the MC
  simulation is changed within limits deduced from the measured 
$x_\gamma$ distribution~\cite{krzys}, leading to $\pm 1$\% systematic error due to a different acceptance.

\item The luminosity measurement has an error of  $3.4$\%.

\end{itemize}

The effects of each systematic error
on the cross sections are determined by evaluating an alternative unfolding matrix {\bf A$^\prime$} using  
the MC prediction made with the corresponding systematic variation applied.
The differences to the default unfolding matrix {\bf A$^\prime$-A} are used to evaluate the contributions to
the error matrices of the results
using standard error propagation. 
The final error matrix is split into fully correlated and fully uncorrelated parts which are listed in 
tables~\ref{tab:inclusive} to~\ref{tab:pperp2}.
The  systematic  uncertainty obtained on the total inclusive cross section is 
 $\pm 13$\%.
The largest contribution to this uncertainty arises
from the systematic uncertainties attributed to the description of the shower shapes.

\section{Results} 
\label{sect:res}

The prompt photon cross sections presented below are given for the phase space
defined in table~\ref{tab:kinem}.

\renewcommand{\arraystretch}{1.2} 
\begin{table}[tt]
\begin{center}
\begin{tabular}{|c|c|}
\hline
\multicolumn{2}{|c|}{\bf \boldmath{H1 Prompt Photon Phase Space} }\\
\hline
\multirow{5}{30mm}{\bf  ~~~~~~~Inclusive  \\ ~~~~cross section} & $6 < \etg < 15$ GeV \\
   &  $-1.0 < \etag < 2.4$ \\
   &  $z=E_T^\gamma/E_T^{\gamma\mathrm{-jet}} > 0.9$ \\
   &  $\qsq < 1$ GeV$^2$ \\
   &  $0.1< y <0.7$ \\ \hline
\multirow{2}{30mm}{\bf  ~~~~Jet definition} &  $\Etj > 4.5 $ GeV\\
               & $-1.3 < \etajet < 2.3$ \\  \hline
\end{tabular}
\caption{ Phase space for the measurement of prompt photon  cross
sections. Kinematics are defined in the H1 laboratory frame.}
\label{tab:kinem}
\end{center}
\end{table}

Bin averaged differential cross sections are presented
in figures~\ref{fig:xsec_incl} to~\ref{fig:xsec_pperp} and in  tables~\ref{tab:inclusive} to~\ref{tab:pperp2}.
For all measurements the total uncertainty is dominated by the systematic errors. 
    The figures also show the ratio 
     of the NLO QCD prediction (FGH)~\cite{heinrich,heinrich2} to the measured cross section
       $R = \sigma_{FGH}/\sigma_{meas}$ with
        the uncertainty of the NLO calculation.
The factors $f_{corr}$ (see section~\ref{sect:theo}) for the correction of the theoretical calculations  for hadronisation, multi parton interactions and the definition of the isolation are given in the cross section tables with their errors.
 
The measured inclusive prompt photon cross section  in the phase space defined in table~\ref{tab:kinem} is
\begin{displaymath}
\sigma(ep\rightarrow e \gamma X) = 66.9 \pm 1.8 \;\mbox{(stat)}\;\pm 8.7 \;\mbox{(syst)}\ \mbox{pb.}
\end{displaymath}
Both calculations predict   lower cross sections of 
$52.1^{\,+\,5.3}_{\,-\,3.4}$~pb (FGH) and $56.7^{\,+\,2.3}_{\,-\,3.1}$~pb (LZ), while the
 MC expectation from PYTHIA is $46.4$~pb.
Theoretical uncertainties due to missing higher orders are estimated by simultaneously varying $\mu_R$ and $\mu_F$ by a factor of $0.5$ to $2.0$.
In addition, the errors on the theoretical predictions include  uncertainties due to the error of $f_{corr}$ and due to  the PDFs. 
All these error sources are added in quadrature.

\par
Differential inclusive prompt photon cross sections $\mbox{d}\sigma/\mbox{d}\etg$ and $\mbox{d}\sigma/\mbox{d}\etag$ are presented in 
table~\ref{tab:inclusive} and in figure~\ref{fig:xsec_incl}.
The results are compared to a QCD   calculation based on the collinear factorisation in 
NLO (FGH)~\cite{heinrich,heinrich2}, to a calculation based on the $k_T$ factorisation approach (LZ)~\cite{zotov}.
Both calculations are below the data, most significantly at low $\etg$.
The LZ calculation gives a reasonable description of the shape of $\etag$, whereas the FGH calculation is  significantly 
below the data for central and backward photons ($\etag<0.9$).

Double differential cross sections $d^2\sigma/\mbox{d}\etg \mbox{d}\etag$ are shown in 
figure~\ref{fig:xsec_incl2}  and table~\ref{tab:incl2}  for all five bins in $\etag$. 
The $\etag$ bins correspond to the wheel structure of the LAr calorimeter.
LZ provides a reasonable description of the data with the exception of the lowest $\etg$ bin  in the central $\etag$ ($0.2<\etag<0.9$) region.
The FGH calculation underestimates the cross section in the central ($0.2<\etag<0.9$) and backward ($\etag<-0.6$) region. Here,
it is significantly below the data. 
The prediction from PYTHIA is also shown.
It underestimates the measured cross section by roughly $45\%$, most significantly at low $\Etg$.

The  prompt photon plus jet cross section  is
\begin{displaymath}
\sigma(ep\rightarrow e \gamma \mathrm{~jet~} X) = 50.1 \pm 1.7 \;\mbox{(stat)}\;\pm 6.5 \; \mbox{(syst)}\ \mbox{pb.}
\end{displaymath}
It is similar to the inclusive cross section, since the prompt
photon recoils most of the time against a prominent hadronic jet.
The theoretical calculations predict  cross sections of $40.6^{\,+\,5.3}_{\,-\,1.9}$~pb (FGH) and $45.7^{\,+\,4.7}_{\,-\,2.1}$~pb (LZ).  Both are 
compatible with the measurement within the errors. The PYTHIA expectation of  $33.9$~pb  is again too low.

Cross sections for the production of a prompt photon plus jet are presented in figure~\ref{fig:xsec_excl} and tables~\ref{tab:exclusive},~\ref{tab:xgamma}
 as a function of the variables $\etg$, $\etag$, $\etjet$ and $\etajet$.
Both calculations 
 give a reasonable description of the $\etg$ and $\etjet$ cross sections but show deficits in the description of the $\etaj$ shape.
Here, the LZ prediction  is  too high for jets with $\etaj<0.5$, and
both calculations underestimate the rate of events with forward jets. 
As in the inclusive case, the FGH prediction is too low for $\etag<0.2$.

Photon plus jet cross section as a function of 
the estimators $\xglo$ and $\xplo$ are shown in figure~\ref{fig:xsec_xg} and table~\ref{tab:exclusive}. 
Both distributions are described  by the calculations within errors.

Cross sections for the two observables describing the transverse 
correlation between the photon and the jet, $\pperp$ and $\Dphi$,  are shown in figure~\ref{fig:xsec_pperp} and tables~\ref{tab:pperp1},~\ref{tab:pperp2}.
Both variables  are expected to be sensitive  to higher order gluon emission.
The phase-space is divided into two parts: one with  $\xglo>0.8$ where the direct interaction
of a photon with the proton dominates and one with  $\xglo<0.8$, including  
significant contributions from events with a resolved photon.
For $\xglo>0.8$  both predictions underestimate the tails of the distributions suggesting that there is 
more decorrelation in the data than predicted.
For $\xglo<0.8$ the
 $\pperp$ distribution is harder than for $\xglo>0.8$, 
which reflects the increased contributions from events with a resolved 
photon and from photons radiated from quarks in di-jet events. 
The FGH calculation poorly describes the $\pperp$ distribution but gives a reasonable 
description of  the measurement in $\Dphi$ for $\xglo<0.8$, 
except for the highest bin in $\Dphi$.
The regions $\Dphi\rightarrow 180^\circ$ and $\pperp\rightarrow 0$ are sensitive to multiple soft gluon radiation which limits the validity of  fixed order calculations~\cite{gudrun}.
The LZ calculation includes multiple  gluon radiation in the initial state
before the hard subprocess and describes $\Dphi>170^\circ$ and $\pperp <2$~GeV,  but 
predicts a significantly lower contribution of events in the tails of
 both distributions as compared to the data.

The present measurement is compared to the published results of H1~\cite{H1gamma} and ZEUS~\cite{ZEUS2} in the restricted phase space $0.2 < y_h < 0.7$.
For the comparison with the inclusive measurement of H1 the $\etag$ range is  restricted to  $-1.0 <\etag < 0.9$. 
For the comparison with the ZEUS results for isolated photons with a jet,
 the kinematic range is changed to $7< \etg< 15 $~GeV, $6 < \etj < 17$~GeV and $-1.6 < \etaj<2.4$. 
The results of this analysis are found in agreement with the previous measurements~\cite{krzys}.

\section{Conclusions} 

The photoproduction of prompt  photons  is measured
in $ep$ collisions at a centre-of-mass energy of $319~{\rm GeV}$ with
the H1 detector  at HERA using a data sample corresponding to an integrated
 luminosity of $340~{\rm pb}^{-1}$.
Photons with a transverse energy in the range  $6 < \etg < 15$~GeV
 and with pseudorapidity   $-1.0 < \etag < 2.4$
are measured in the kinematic region  $Q^2<1$~GeV$^2$ and $0.1<y <0.7$. 
Compared to previous measurements, the range of $\etag$  is significantly extended, and the luminosity of the measurement is increased by a factor three.

\par
Single differential and double differential cross sections are measured.
The data are compared to a QCD   calculation based on the collinear factorisation in 
NLO (FGH)~\cite{heinrich,heinrich2}, to a QCD calculation based on the $k_T$ factorisation approach (LZ)~\cite{zotov}, and to the MC prediction from PYTHIA.
The predicted total cross section is lower than the measurement by
around $20\%$.
Both theoretical calculations underestimate the data at low $\etg$.
 While the LZ prediction describes the shape
of $\mbox{d}\sigma/\mbox{d}\eta^\gamma$ reasonably well, the FGH prediction is
significantly below the data for  backward photons ($\etag<-0.6$).
PYTHIA underestimates the data by roughly $45\%$, most significantly at low $\etg$.

\par
 Differential cross sections for photon plus jet are measured as 
a function of the observables $\etg$, $\etag$, $\etaj$, $\etj$, $\xglo$, and $\xplo$.
The measured cross sections as a function of the transverse energy of the photon and the jet as well as
  $\xglo$ and $\xplo$ are
described within errors by the calculations.
However, neither of the predictions is able to describe the measured shape as a function of $\etaj$.

Correlations in the transverse plane between the jet and the photon are investigated by measurements of the difference in azimuthal angle
$\Dphi$ and of the photon's  momentum perpendicular to the jet direction, $\pperp$.
A significant fraction of events shows a topology which is not back-to-back. 
Neither calculation is able to describe the measured correlations in the transverse plane.

Prompt photon cross section in photoproduction are now measured at a precision of about $10\%$, with hadronisation corrections known at the level of $5\%$.
The challenge remains to further improve the theoretical calculations and arrive at a deeper understanding of the underlying QCD dynamics in this interesting channel.

\section*{Acknowledgements}

We are grateful to the HERA machine group whose outstanding
efforts have made this experiment possible. 
We thank the engineers and technicians for their work in constructing and
maintaining the H1 detector, our funding agencies for 
financial support, the
DESY technical staff for continual assistance
and the DESY directorate for support and for the
hospitality which they extend to the non DESY 
members of the collaboration.
We would like to thank Artem Lipatov and Nikolai Zotov for providing the LZ calculations and Gudrun Heinrich for help with the FGH calculations.

\begin{appendix} 
\section{Unfolding procedure}
\label{sect:unfold}

The photon signal is extracted using an unfolding procedure to relate
distributions $\vec{y}_{\mathrm{rec}}$ of reconstructed variables 
to distributions $\vec{x}_{\mathrm{true}}$ of 
true variables on hadron level, to determine the fractions of signal 
and background and to correct the data for the detector efficiency.
The unfolding  matrix \matrixA{} which reflects the acceptance of the H1 detector relates the two vectors,  
$\matrixA \vec{x}_{\mathrm{true}} =\vec{y}_{\mathrm{rec}} $.
Each matrix element $A_{ij}$ is the probability for an event originating from bin $j$ of $\vec{x}_{\mathrm{true}}$ to be measured in 
bin $i$ of $\vec{y}_{\mathrm{rec}}$. The matrix \matrixA{} is computed using the PYTHIA simulation for the signal and the background, interfaced to the GEANT simulation of the H1 detector.

A schematic view of the simplified unfolding matrix \matrixA{} is shown in figure~\ref{fig:unfold}. 
Each row of the  matrix corresponds to one element of the vector $\vec{x}_{\mathrm{true}}$. The elements of $\vec{x}_{\mathrm{true}}$ are: signal, migration and background bins.
Each column of the matrix corresponds to one element of the vector $\vec{y}_{\mathrm{rec}}$. The elements of  $\vec{y}_{\mathrm{rec}}$ are: reconstructed bins and side bins.
 When solving the equation for $\vec{x}_{\mathrm{true}}$ the number of efficiency corrected signal, migration and background events is determined in one step.

\begin{figure}[Ht]
\setcounter{figure}{9}
  \begin{center}
    \includegraphics[width=1.0\textwidth]{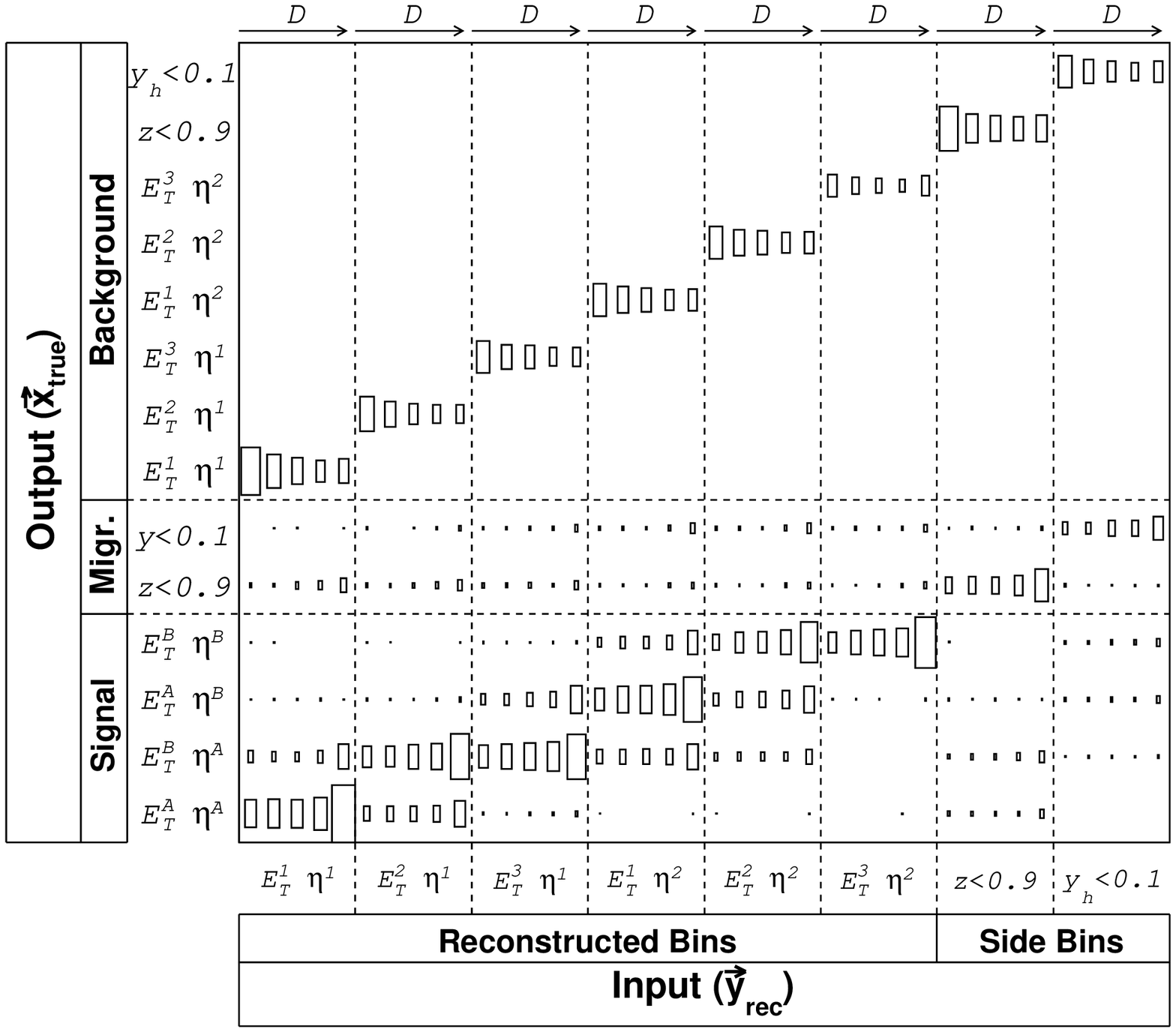}
    \caption{
Schematic sketch of the unfolding matrix for $2\times2$ signal bins and $3\times2\times5$ reconstructed bins. 
Two generator cuts ($z$ and $y$) and the background are taken into account.
The size of the boxes reflects the number of entries in a bin.}
    \label{fig:unfold}
  \end{center}
\end{figure}

The input $\vec{y}_{\mathrm{rec}}$ is binned in three dimensions in the reconstructed quantities $E_T^{1,2,3}$, $\eta^{1,2}$ and $D$. The binning in $D$ is required for the discrimination of signal and background.
Figure~\ref{fig:unfold} shows  $3\times2\times5$ ``Reconstructed Bins''.
The signal is binned in the hadron-level quantities $E_T^{A,B}$ and $\eta^{A,B}$. Figure~\ref{fig:unfold} shows $2\times 2$ ``Signal'' bins in these variables.

In addition, $\vec{x}_{\mathrm{true}}$ includes ``background'' bins in $E_T^{1,2,3}$ and $\eta^{1,2}$, in parallel to the reconstructed quantities. These bins give the amount of background in each reconstructed bin. The background is determined in the unfolding together with the signal contribution.

The final unfolding matrix \matrixA{} also takes into account migrations into or out of the phase space of the measurement. 
For each cut on hadron level, used to define the measurement phase space (table~\ref{tab:kinem}),
a migration bin is added, containing events generated outside of the phase space but reconstructed in any of the input bins.
In figure~\ref{fig:unfold}, two such ``Migr.'' bins are shown.
In order to minimise possible biases introduced by the signal MC simulation outside the phase space, each migration bin is subdivided into 
\etg{} and \etag{} bins (not shown in the figure). 

The amount of migration from outside of the generated phase space is controlled by including ``Side'' bins on detector level for each of the ``Migration'' bins on hadron level.
A side bin is defined as a narrow slice outside the nominal cut value of the reconstructed variable. 
The side bins are also subdivided into $\etg$ and $\etag$ bins.

Using matrix  \matrixA{} the unfolded distribution $\vec{x}_{\mathrm{true}}$ is obtained from the observed distribution  $\vec{y}_{\mathrm{rec}}$ 
by minimising a $\chi^2$ function given by 
\beq
\chi^{2}=\chi^{2}_{A} + \tau^2 \chi^{2}_{L}, 
\eeq
where
\beq
\chi^{2}_{A}=1/2 \cdot (\vec{y}_\mathrm{rec}-\matrixA\vec{x}_\mathrm{true})^{T} \textbf{V}^{-1} (\vec{y}_\mathrm{rec}-\matrixA\vec{x}_\mathrm{true})
\eeq
measures the deviation of $\matrixA\vec{x}_\mathrm{true}$ from the data bins $\vec{y}_{rec}$.
Here, $\textbf{V} = \mbox{Cov} (y_i, y_j)$ is the covariance matrix of the data, initially approximated by the observed statistical errors.
In order to avoid a known bias of this procedure~\cite{agostini}, the unfolding is iterated using an updated covariance matrix~\cite{krzys}, constructed from the expected statistical uncertainties.
For a given regularisation parameter $\tau$ the regularisation term is defined as $\chi^{2}_{L}=(\vec{x}_\mathrm{true})^{2}$.
The minimum $\chi^{2}$ can be  calculated analytically and is found as
\beq
 \vec{x}_\mathrm{true}=((\matrixA^{T} \textbf{V}^{-1} \matrixA)+{\bf 1} \tau)^{-1} \matrixA^{T} \textbf{V}^{-1} \vec{y}_\mathrm{rec}.
\eeq
The size of the regularisation parameter $\tau$ is chosen using the $L$-curve method~\cite{lcurve1,lcurve2,lcurve3}.

 \end{appendix}

\begin{flushleft}

\end{flushleft}

\clearpage


\renewcommand{\arraystretch}{1.2} 
\clearpage

\vspace*{\fill}
\begin{table}[ht]
\begin{center}
    \begin{tabular}{|r @{,} l|rll||c|}
      \hline
      \multicolumn{6}{|c|}{\bf \boldmath H1 Inclusive Prompt Photon Cross Sections }  \\
      \hline
      \hline
      \multicolumn{2}{|c|}{$\etg$} & $ \mbox{d}\sigma /\mbox{d} \etg$ & uncorr. & corr. & $f_{corr}$  \\
      \multicolumn{2}{|c|}{[GeV]} & \multicolumn{3}{c||}{[pb/GeV]} &  \\
      \hline

 $ [ 6.0 $&$ 7.0 ] $ & $ 27.24 $ & $ \pm  1.86 $ & $ \pm 3.04 $ & $0.88^{\,+\,0.02}_{\,-\,0.04}$\\
 $ [ 7.0 $&$ 8.5 ] $ & $ 12.94 $ & $ \pm  0.71 $ & $ \pm 1.94 $ & $0.89^{\,+\,0.02}_{\,-\,0.04}$\\
 $ [ 8.5 $&$ 10.0 ] $ & $ 6.74 $ & $ \pm  0.65 $ & $ \pm 0.95 $ & $0.93^{\,+\,0.02}_{\,-\,0.04}$\\
 $ [ 10.0 $&$ 15.0 ] $ & $ 2.02 $ & $ \pm  0.17 $ & $ \pm 0.24 $ & $0.96^{\,+\,0.02}_{\,-\,0.03}$\\

      \hline
      \hline
      \multicolumn{2}{|c|}{\etag} & $ \mbox{d}\sigma /\mbox{d} \etag$ & uncorr. & corr. & $f_{corr}$  \\
      \multicolumn{2}{|c|}{} & \multicolumn{3}{c||}{[pb]} &  \\
      \hline

 $ [ -1.00 $&$ -0.57 ] $ & $ 18.4 $ & $ \pm  1.3 $ & $ \pm 2.5 $ & $0.99^{\,+\,0.02}_{\,-\,0.04}$\\
 $ [ -0.57 $&$ 0.20 ] $ & $ 23.9 $ & $ \pm  1.5 $ & $ \pm 1.5 $ & $0.94^{\,+\,0.02}_{\,-\,0.03}$\\
 $ [ 0.20 $&$ 0.94 ] $ & $ 27.7 $ & $ \pm  1.2 $ & $ \pm 2.3 $ & $0.90^{\,+\,0.01}_{\,-\,0.04}$\\
 $ [ 0.94 $&$ 1.42 ] $ & $ 19.3 $ & $ \pm  1.3 $ & $ \pm 3.0 $ & $0.87^{\,+\,0.03}_{\,-\,0.04}$\\
 $ [ 1.42 $&$ 2.40 ] $ & $ 11.0 $ & $ \pm  1.0 $ & $ \pm 3.5 $ & $0.84^{\,+\,0.02}_{\,-\,0.05}$\\

      \hline
    \end{tabular}
    \caption{Bin averaged differential cross sections for inclusive prompt photon production as a function of $\etag$ and $\etg$
in the kinematic range specified in table~\ref{tab:kinem}.
The bin ranges, the differential cross section  values, the uncorrelated and correlated uncertainties, and $f_{corr}$ are listed, where
$f_{corr}$ denotes the correction factor applied to the theoretical cross sections. It corrects for multi parton interactions, hadronisation and the different 
algorithm used for the isolation cut.
}
    \label{tab:inclusive}
  \end{center}
\end{table}
\vspace*{\fill}
\vspace*{\fill}
\begin{table}[ht]
  \begin{center}
    \begin{tabular}{|r @{,} l| r @{,} l|rll||c|}
      \hline
      \multicolumn{8}{|c|}{\bf \boldmath H1 Inclusive Prompt Photon Cross Sections }  \\
      \hline
  \multicolumn{2}{|c|}{\etag} &\multicolumn{2}{|c|}{\etg} & $ \mbox{d}^2\sigma /\mbox{d} \etg \mbox{d} \etag$ & uncorr. & corr. & $f_{corr}$ \\
      \multicolumn{2}{|c|}{} &\multicolumn{2}{|c|}{[GeV]} & \multicolumn{3}{c||}{[pb/GeV]} &  \\

      \hline
          \hline
$[-1.00 $&$-0.57]$
 &  $ [ 6.00 $&$ 7.00 ] $ & $ 9.24 $ & $ \pm  0.93 $ & $ \pm 1.06 $ & $0.99^{\,+\,0.02}_{\,-\,0.04}$\\
 \multicolumn{2}{|c|}{} &  $ [ 7.00 $&$ 8.50 ] $ & $ 3.75 $ & $ \pm  0.57 $ & $ \pm 0.44 $ & $0.99^{\,+\,0.02}_{\,-\,0.05}$\\
 \multicolumn{2}{|c|}{} &  $ [ 8.50 $&$ 10.00 ] $ & $ 1.43 $ & $ \pm  0.41 $ & $ \pm 0.24 $ & $0.99^{\,+\,0.02}_{\,-\,0.02}$\\
 \multicolumn{2}{|c|}{} &  $ [ 10.00 $&$ 15.00 ] $ & $ 0.27 $ & $ \pm  0.10 $ & $ \pm 0.07 $ & $1.01^{\,+\,0.01}_{\,-\,0.03}$\\
      \hline
      \hline
$[-0.57 $&$ 0.20]$
 &  $ [ 6.00 $&$ 7.00 ] $ & $ 9.19 $ & $ \pm  1.04 $ & $ \pm 0.59 $ & $0.93^{\,+\,0.02}_{\,-\,0.04}$\\
  \multicolumn{2}{|c|}{}&  $ [ 7.00 $&$ 8.50 ] $ & $ 5.02 $ & $ \pm  0.59 $ & $ \pm 0.48 $ & $0.94^{\,+\,0.02}_{\,-\,0.02}$\\
  \multicolumn{2}{|c|}{}&  $ [ 8.50 $&$ 10.00 ] $ & $ 2.29 $ & $ \pm  0.46 $ & $ \pm 0.19 $ & $0.96^{\,+\,0.02}_{\,-\,0.03}$\\
  \multicolumn{2}{|c|}{}&  $ [ 10.00 $&$ 15.00 ] $ & $ 0.76 $ & $ \pm  0.12 $ & $ \pm 0.06 $ & $1.00^{\,+\,0.02}_{\,-\,0.03}$\\
      \hline 
      \hline
$[0.20 $&$ 0.94]$
  &  $ [ 6.00 $&$ 7.00 ] $ & $ 10.90 $ & $ \pm  0.86 $ & $ \pm 0.78 $ & $0.87^{\,+\,0.01}_{\,-\,0.05}$\\
  \multicolumn{2}{|c|}{} &  $ [ 7.00 $&$ 8.50 ] $ & $ 5.15 $ & $ \pm  0.43 $ & $ \pm 0.52 $ & $0.89^{\,+\,0.02}_{\,-\,0.03}$\\
  \multicolumn{2}{|c|}{} &  $ [ 8.50 $&$ 10.00 ] $ & $ 3.28 $ & $ \pm  0.34 $ & $ \pm 0.24 $ & $0.94^{\,+\,0.01}_{\,-\,0.04}$\\
   \multicolumn{2}{|c|}{}&  $ [ 10.00 $&$ 15.00 ] $ & $ 0.83 $ & $ \pm  0.10 $ & $ \pm 0.07 $ & $0.96^{\,+\,0.02}_{\,-\,0.04}$\\
      \hline 
      \hline
$[0.94 $ & $ 1.42]$
  &  $ [ 6.00 $&$ 7.00 ] $ & $ 7.68 $ & $ \pm  1.02 $ & $ \pm 1.18 $ & $0.83^{\,+\,0.04}_{\,-\,0.04}$\\
  \multicolumn{2}{|c|}{} &  $ [ 7.00 $&$ 8.50 ] $ & $ 3.31 $ & $ \pm  0.51 $ & $ \pm 0.51 $ & $0.85^{\,+\,0.02}_{\,-\,0.05}$\\
  \multicolumn{2}{|c|}{} &  $ [ 8.50 $&$ 10.00 ] $ & $ 2.27 $ & $ \pm  0.44 $ & $ \pm 0.38 $ & $0.90^{\,+\,0.03}_{\,-\,0.04}$\\
   \multicolumn{2}{|c|}{}&  $ [ 10.00 $&$ 15.00 ] $ & $ 0.66 $ & $ \pm  0.12 $ & $ \pm 0.09 $ & $0.95^{\,+\,0.03}_{\,-\,0.03}$\\
      \hline 
      \hline
$[1.42 $&$ 2.43]$
  &  $ [ 6.00 $&$ 7.00 ] $ & $ 4.54 $ & $ \pm  0.90 $ & $ \pm 1.29 $ & $0.79^{\,+\,0.02}_{\,-\,0.05}$\\
  \multicolumn{2}{|c|}{}&  $ [ 7.00 $&$ 8.50 ] $ & $ 2.12 $ & $ \pm  0.41 $ & $ \pm 0.72 $ & $0.82^{\,+\,0.02}_{\,-\,0.05}$\\
  \multicolumn{2}{|c|}{}&  $ [ 8.50 $&$ 10.00 ] $ & $ 0.86 $ & $ \pm  0.31 $ & $ \pm 0.35 $ & $0.88^{\,+\,0.02}_{\,-\,0.06}$\\
   \multicolumn{2}{|c|}{}&  $ [ 10.00 $&$ 15.00 ] $ & $ 0.40 $ & $ \pm  0.08 $ & $ \pm 0.09 $ & $0.93^{\,+\,0.03}_{\,-\,0.03}$\\
      \hline 

      \hline
    \end{tabular}
    \caption{Bin averaged double differential cross section for
      inclusive prompt photon production in   
bins of transverse energy and pseudorapidity of the photon.
The errors are correlated between different $\etg$ and different $\etag$ bins.
More details are given in the caption of table~\ref{tab:inclusive}.}
    \label{tab:incl2}
  \end{center}
\end{table}
\vspace*{\fill}
\vspace*{\fill}
\begin{table}[hht]
  \begin{center}
    \begin{tabular}{|r @{,} l|rll||c|}
      \hline
      \multicolumn{6}{|c|}{\bf \boldmath H1 Prompt Photon plus Jet Cross Sections}  \\
      \hline
      \hline
       \multicolumn{2}{|c|}{$\etg$} & $ \mbox{d}\sigma /\mbox{d} \etg $ & uncorr. & corr. & $f_{corr}$  \\
      \multicolumn{2}{|c|}{[GeV]} & \multicolumn{3}{c||}{[pb/GeV]} & \\
      \hline

 $ [ 6.0 $&$ 7.0 ] $ & $ 18.53 $ & $ \pm  1.61 $ & $ \pm 2.21 $ & $0.82^{\,+\,0.05}_{\,-\,0.06}$\\
 $ [ 7.0 $&$ 8.5 ] $ & $ 9.93 $ & $ \pm  1.06 $ & $ \pm 1.39 $ & $0.87^{\,+\,0.04}_{\,-\,0.05}$\\
 $ [ 8.5 $&$ 10.0 ] $ & $ 5.50 $ & $ \pm  0.45 $ & $ \pm 0.73 $ & $0.92^{\,+\,0.03}_{\,-\,0.04}$\\
 $ [ 10.0 $&$ 15.0 ] $ & $ 1.68 $ & $ \pm  0.23 $ & $ \pm 0.16 $ & $0.95^{\,+\,0.03}_{\,-\,0.03}$\\
      \hline
      \hline
       \multicolumn{2}{|c|}{\etag} & $ \mbox{d}\sigma /\mbox{d} \etag$ & uncorr. & corr. & $f_{corr}$ \\
      \multicolumn{2}{|c|}{} & \multicolumn{3}{c||}{[pb]} &\\
      \hline

 $ [ -1.00 $&$ -0.57 ] $ & $ 14.79 $ & $ \pm  1.25 $ & $ \pm 1.70 $ & $0.94^{\,+\,0.06}_{\,-\,0.07}$\\
 $ [ -0.57 $&$ 0.20 ] $ & $ 18.57 $ & $ \pm  1.47 $ & $ \pm 1.75 $ & $0.90^{\,+\,0.04}_{\,-\,0.05}$\\
 $ [ 0.20 $&$ 0.94 ] $ & $ 21.12 $ & $ \pm  1.21 $ & $ \pm 1.77 $ & $0.87^{\,+\,0.03}_{\,-\,0.04}$\\
 $ [ 0.94 $&$ 1.42 ] $ & $ 13.88 $ & $ \pm  3.15 $ & $ \pm 2.12 $ & $0.86^{\,+\,0.04}_{\,-\,0.04}$\\
 $ [ 1.42 $&$ 2.40 ] $ & $ 7.31 $ & $ \pm  2.73 $ & $ \pm 1.54 $ & $0.84^{\,+\,0.03}_{\,-\,0.05}$\\
      \hline
      \hline
     \multicolumn{2}{|c|}{\etj} & $ \mbox{d}\sigma /\mbox{d} \etj$ & uncorr. & corr. & $f_{corr}$ \\
     \multicolumn{2}{|c|}{[GeV]} & \multicolumn{3}{c||}{[pb/GeV]} & \\
      \hline

 $ [ 4.5 $&$ 6.2 ] $ & $ 6.60 $ & $ \pm  0.74 $ & $ \pm 1.42 $ & $0.85^{\,+\,0.04}_{\,-\,0.05}$\\
 $ [ 6.2 $&$ 8.0 ] $ & $ 6.93 $ & $ \pm  1.08 $ & $ \pm 0.84 $ & $0.83^{\,+\,0.05}_{\,-\,0.06}$\\
 $ [ 8.0 $&$ 10.0 ] $ & $ 6.15 $ & $ \pm  0.78 $ & $ \pm 0.65 $ & $0.90^{\,+\,0.03}_{\,-\,0.05}$\\
 $ [ 10.0 $&$ 15.0 ] $ & $ 1.88 $ & $ \pm  0.33 $ & $ \pm 0.17 $ & $0.96^{\,+\,0.03}_{\,-\,0.04}$\\
      \hline
      \hline
     \multicolumn{2}{|c|}{\etaj} & $ \mbox{d}\sigma /\mbox{d} \etaj$ & uncorr. & corr. & $f_{corr}$\\
      \multicolumn{2}{|c|}{} & \multicolumn{3}{c||}{[pb]} & \\
      \hline

 $ [ -1.3 $&$ -0.4 ] $ & $ 7.1 $ & $ \pm  0.7 $ & $ \pm 1.0 $ & $0.80^{\,+\,0.03}_{\,-\,0.04}$\\
 $ [ -0.4 $&$ 0.5 ] $ & $ 14.9 $ & $ \pm  0.8 $ & $ \pm 2.1 $ & $0.86^{\,+\,0.03}_{\,-\,0.04}$\\
 $ [ 0.5 $&$ 1.4 ] $ & $ 18.7 $ & $ \pm  1.1 $ & $ \pm 2.7 $ & $0.91^{\,+\,0.04}_{\,-\,0.06}$\\
 $ [ 1.4 $&$ 2.3 ] $ & $ 15.3 $ & $ \pm  1.2 $ & $ \pm 2.2 $ & $0.94^{\,+\,0.04}_{\,-\,0.06}$\\
      \hline
    \end{tabular}
    \caption
    {Bin averaged differential cross section for prompt photon plus jet production as a function of $\etg$, $\etag$, $\etj$ and $\etaj$.
More details are given in the caption of table~\ref{tab:inclusive}.}
    \label{tab:exclusive}
  \end{center}
\end{table}
\vspace*{\fill}
\clearpage
\vspace*{\fill}
\begin{table}[hht]
  \begin{center}
    \begin{tabular}{|r @{,} l|rll||c|}
      \hline
      \multicolumn{6}{|c|}{\bf \boldmath H1 Prompt Photon plus Jet Cross Sections }  \\
      \hline
      \hline
      \multicolumn{2}{|c|}{\xglo} & $ \mbox{d}\sigma /\mbox{d} \xglo$ & uncorr. & corr. & $f_{corr}$\\
     \multicolumn{2}{|c|}{} & \multicolumn{3}{c||}{[pb]} & \\
      \hline

 $ [ 0.0 $&$ 0.5 ] $ & $ 23.0 $ & $ \pm  3.3 $ & $ \pm 3.8 $ & $0.78^{\,+\,0.04}_{\,-\,0.08}$\\
 $ [ 0.5 $&$ 0.7 ] $ & $ 44.1 $ & $ \pm  9.6 $ & $ \pm 8.1 $ & $0.89^{\,+\,0.06}_{\,-\,0.08}$\\
 $ [ 0.7 $&$ 0.9 ] $ & $ 70.3 $ & $ \pm  12.3 $ & $ \pm 10.9 $ & $1.24^{\,+\,0.04}_{\,-\,0.06}$\\
 $ [ 0.9 $&$ 1.1 ] $ & $ 75.9 $ & $ \pm  9.8 $ & $ \pm 8.4 $ & $0.82^{\,+\,0.04}_{\,-\,0.04}$\\
      \hline
      \hline
       \multicolumn{2}{|c|}{\xplo} & $ \mbox{d}\sigma /\mbox{d} \xplo$ & uncorr. & corr. & $f_{corr}$ \\
     \multicolumn{2}{|c|}{} & \multicolumn{3}{c||}{[pb]} & \\
      \hline

 $ [ 0.001 $&$ 0.010 ] $ & $ 1257 $ & $ \pm  91 $ & $ \pm 102 $ & $0.84^{\,+\,0.04}_{\,-\,0.05}$\\
 $ [ 0.010 $&$ 0.025 ] $ & $ 1325 $ & $ \pm  65 $ & $ \pm 150 $ & $0.88^{\,+\,0.04}_{\,-\,0.05}$\\
 $ [ 0.025 $&$ 0.040 ] $ & $ 698 $ & $ \pm  90 $ & $ \pm 134 $ & $0.90^{\,+\,0.03}_{\,-\,0.05}$\\
 $ [ 0.040 $&$ 0.060 ] $ & $ 341 $ & $ \pm  47 $ & $ \pm 66 $ & $0.91^{\,+\,0.03}_{\,-\,0.06}$\\
  \hline
    \end{tabular}
    \caption
    {Bin averaged differential cross section for prompt photon plus jet production as a function of $\xglo$ and $\xplo$.
 More details are given in the caption of table~\ref{tab:inclusive}.}
    \label{tab:xgamma}
  \end{center}
\end{table}
\vspace*{\fill}
\clearpage
\vspace*{\fill}
\begin{table}[ht]
  \begin{center}
    \begin{tabular}{|r @{,} l|r @{,} l|rll||c|}
      \hline
      \multicolumn{8}{|c|}{\bf \boldmath H1 Prompt Photon plus Jet Cross Sections}  \\
      \hline
      \hline
      \multicolumn{2}{|c|}{\xglo} & \multicolumn{2}{|c|}{$\pperp$} & $ \mbox{d}\sigma /\mbox{d} \pperp$ & uncorr. & corr. & $f_{corr}$ \\
     \multicolumn{2}{|c|}{}& \multicolumn{2}{|c|}{[GeV]} & \multicolumn{3}{c||}{[pb/GeV]} & \\
      \hline
$[0.8$&$1.1]$
  &  $ [ 0 $&$ 2 ] $ & $ 7.75 $ & $ \pm  0.50 $ & $ \pm 0.32 $ & $0.89^{\,+\,0.04}_{\,-\,0.04}$\\
  \multicolumn{2}{|c|}{}&  $ [ 2 $&$ 4 ] $ & $ 3.96 $ & $ \pm  0.43 $ & $ \pm 0.53 $ & $0.87^{\,+\,0.01}_{\,-\,0.01}$\\
  \multicolumn{2}{|c|}{}&  $ [ 4 $&$ 6 ] $ & $ 2.16 $ & $ \pm  0.55 $ & $ \pm 0.56 $ & $0.84^{\,+\,0.07}_{\,-\,0.07}$\\
  \multicolumn{2}{|c|}{}&  $ [ 6 $&$ 8 ] $ & $ 0.60 $ & $ \pm  0.53 $ & $ \pm 0.36 $ & $0.81^{\,+\,0.08}_{\,-\,0.08}$\\
      \hline
      \hline
$[0.0$&$0.8]$
  &  $ [ 0 $&$ 2 ] $ & $ 7.14 $ & $ \pm  0.48 $ & $ \pm 1.04 $ & $0.84^{\,+\,0.08}_{\,-\,0.11}$\\
  \multicolumn{2}{|c|}{}&  $ [ 2 $&$ 4 ] $ & $ 4.52 $ & $ \pm  0.49 $ & $ \pm 0.80 $ & $0.91^{\,+\,0.03}_{\,-\,0.08}$\\
  \multicolumn{2}{|c|}{}&  $ [ 4 $&$ 6 ] $ & $ 2.91 $ & $ \pm  0.40 $ & $ \pm 0.56 $ & $0.96^{\,+\,0.03}_{\,-\,0.04}$\\
  \multicolumn{2}{|c|}{}&  $ [ 6 $&$ 8 ] $ & $ 2.35 $ & $ \pm  0.44 $ & $ \pm 0.37 $ & $1.07^{\,+\,0.09}_{\,-\,0.10}$\\
      \hline
    \end{tabular}
    \caption{Bin averaged differential cross sections for prompt photon plus jet production  as a function of $\pperp$, the photon's momentum perpendicular to the jet direction in the transverse plane, separated into two regions with $\xglo>0.8$ and  $\xglo<0.8$.
More details are given in the caption of table~\ref{tab:inclusive}.}
    \label{tab:pperp1}
  \end{center}
\end{table}
\vspace*{\fill}
\clearpage
\vspace*{\fill}
\begin{table}[ht]
  \begin{center}
    \begin{tabular}{|r @{,} l|r @{,} l|rll||c|}
      \hline
      \multicolumn{8}{|c|}{\bf \boldmath H1 Prompt Photon plus Jet Cross Sections}  \\
      \hline
      \hline
     \multicolumn{2}{|c|}{\xglo} &\multicolumn{2}{|c|} {$\Dphi$} & $ \mbox{d}\sigma /\mbox{d}\Dphi$ & uncorr. & corr. & $f_{corr}$ \\
     \multicolumn{2}{|c|}{}& \multicolumn{2}{|c|}{} & \multicolumn{3}{c||}{[pb]} & \\
      \hline
$[0.8$&$1.1]$
  &  $ [ 130 $&$ 150 ] $ & $ 0.19 $ & $ \pm  0.04 $ & $ \pm 0.01 $ & $0.82^{\,+\,0.03}_{\,-\,0.04}$\\
 \multicolumn{2}{|c|}{}&  $ [ 150 $&$ 165 ] $ & $ 0.46 $ & $ \pm  0.06 $ & $ \pm 0.06 $ & $0.84^{\,+\,0.01}_{\,-\,0.01}$\\
 \multicolumn{2}{|c|}{} &  $ [ 165 $&$ 172 ] $ & $ 0.89 $ & $ \pm  0.14 $ & $ \pm 0.07 $ & $0.93^{\,+\,0.00}_{\,-\,0.01}$\\
  \multicolumn{2}{|c|}{}&  $ [ 172 $&$ 180 ] $ & $ 1.38 $ & $ \pm  0.04 $ & $ \pm 0.17 $ & $0.89^{\,+\,0.05}_{\,-\,0.05}$\\
      \hline
      \hline
$[0.0$&$0.8]$
  &  $ [ 130 $&$ 150 ] $ & $ 0.27 $ & $ \pm  0.03 $ & $ \pm 0.04 $ & $0.94^{\,+\,0.03}_{\,-\,0.05}$\\
  \multicolumn{2}{|c|}{}&  $ [ 150 $&$ 165 ] $ & $ 0.52 $ & $ \pm  0.06 $ & $ \pm 0.08 $ & $0.91^{\,+\,0.04}_{\,-\,0.08}$\\
 \multicolumn{2}{|c|}{} &  $ [ 165 $&$ 172 ] $ & $ 0.91 $ & $ \pm  0.14 $ & $ \pm 0.18 $ & $0.83^{\,+\,0.07}_{\,-\,0.09}$\\
  \multicolumn{2}{|c|}{}&  $ [ 172 $&$ 180 ] $ & $ 1.21 $ & $ \pm  0.11 $ & $ \pm 0.15 $ & $0.84^{\,+\,0.08}_{\,-\,0.12}$\\
      \hline
    \end{tabular}
    \caption{Bin averaged differential cross sections for prompt photon plus jet production  as a function of 
$\Dphi$, the difference in azimuthal angle between the photon and the jet, separated into two regions with $\xglo>0.8$ and $\xglo<0.8$.
More details are given in the caption of table~\ref{tab:inclusive}.}
    \label{tab:pperp2}
  \end{center}
\end{table}
\vspace*{\fill}
\clearpage

\setcounter{figure}{1}
\vspace*{\fill}
\begin{figure}[Ht]
  \begin{center} 
 \includegraphics[width=1.0\textwidth]{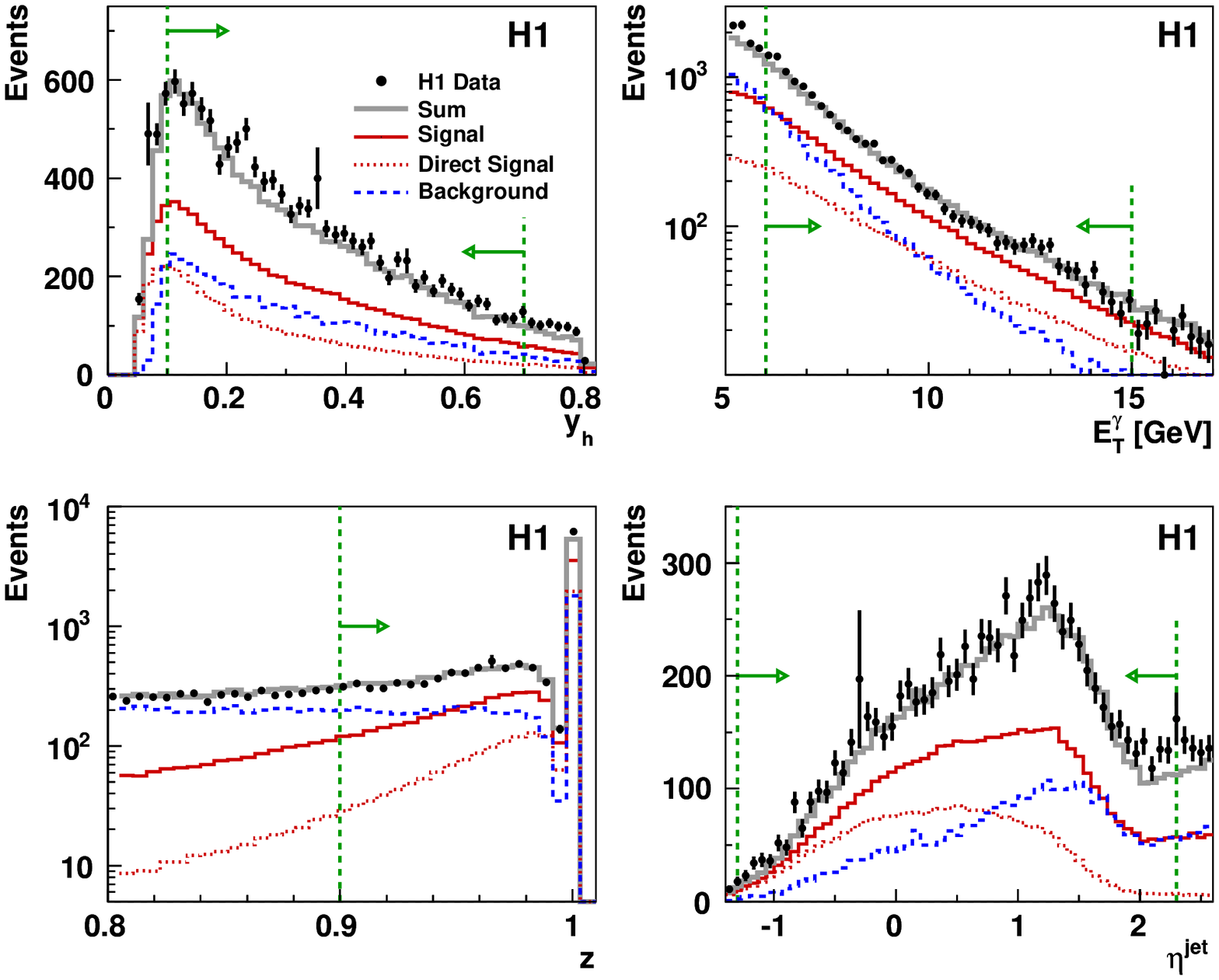}
    \caption{ 
Distributions of a) the inelasticity $y_h$, b) the transverse energy  $E_T^\gamma$ of the photon,  c) the isolation parameter $z$ and d)  $\etaj$ 
for events with prompt  photon candidates.
Data are shown as points with error bars indicating the statistical error.
The  signal prediction from PYTHIA for prompt photons is shown as dark full  line, and the contribution of direct interactions as a dotted line.
The background as simulated with PYTHIA is shown as a dashed line. The signal and background contributions are scaled on average by a factor $1.5$ and $1.7$, respectively.
The sum of the scaled signal and background is shown as the light grey histogram.
The vertical dashed lines indicate the kinematic region of the cross section measurement.}
  \begin{picture} (0.,0.)
     \setlength{\unitlength}{1.0cm}
     \put ( -6.2, 17.2){a)}
     \put ( 2.5 ,17.2){b)}
     \put ( -6.2, 10.7){c)}
     \put (2., 10.7){d)}
  \end{picture}

    \label{fig:control}
  \end{center}
\end{figure}\vspace*{\fill}

\vspace*{\fill}\begin{figure}[Ht]
 \includegraphics[width=1.0\textwidth]{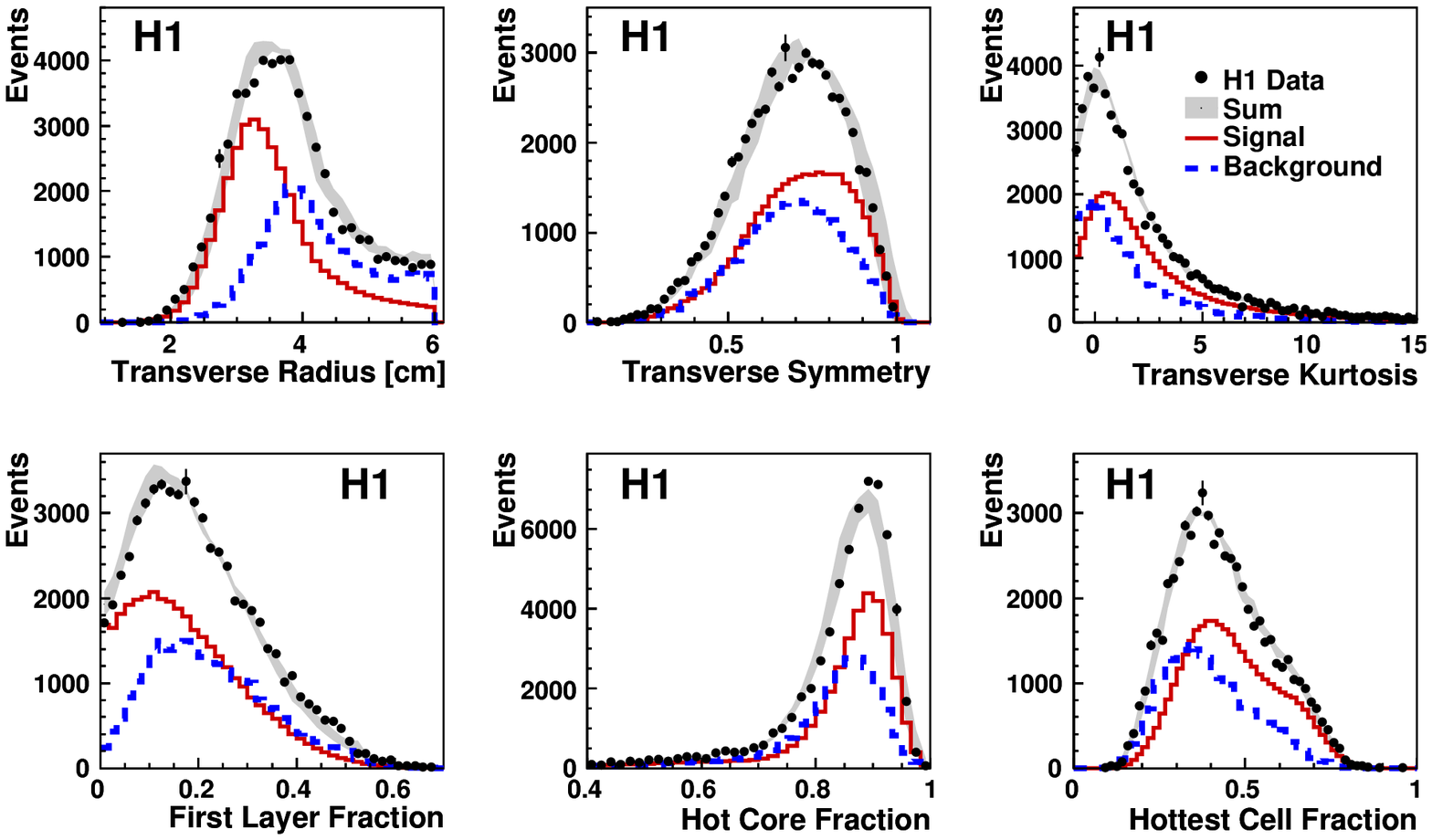}
  \begin{picture} (0.,0.)
     \setlength{\unitlength}{1.0cm}
     \put (1.5, 6.8){a)}
     \put (6.8 ,6.8){b)}
     \put (15.2, 6.8){c)}
     \put (4.5, 1.9){d)}
     \put (6.8, 1.9){e)}
     \put (15.2, 1.9){f)}
  \end{picture}
  \begin{center} 
    \caption{ 
Distributions of the shower shape variables that are used to define the
discriminant for isolated photon identification: 
 a)  transverse radius,
b) transverse symmetry,
 c)  transverse kurtosis,
 d) the  fraction of energy in the first layer,
e) the fraction of energy in the hot core of the cluster and
 f) the fraction of energy in the hottest cell of the cluster
  for all the photon candidates. 
Data are shown as points with error bars for the statistical error.
The expectation from PYTHIA for the signal (background) 
is shown as a full (dashed) line. 
The signal and background contributions are scaled on average by a factor $1.5$ and $1.7$, respectively.
 The shaded 
 band shows the sum of the MC predictions. The width of the band corresponds to  the systematic uncertainty assigned to the description of the shower shapes.
}

    \label{fig:showershapes}
  \end{center}
\end{figure}\vspace*{\fill}

\vspace*{\fill}\begin{figure}[ht]
  \begin{center}
    \includegraphics[width=1.\textwidth]{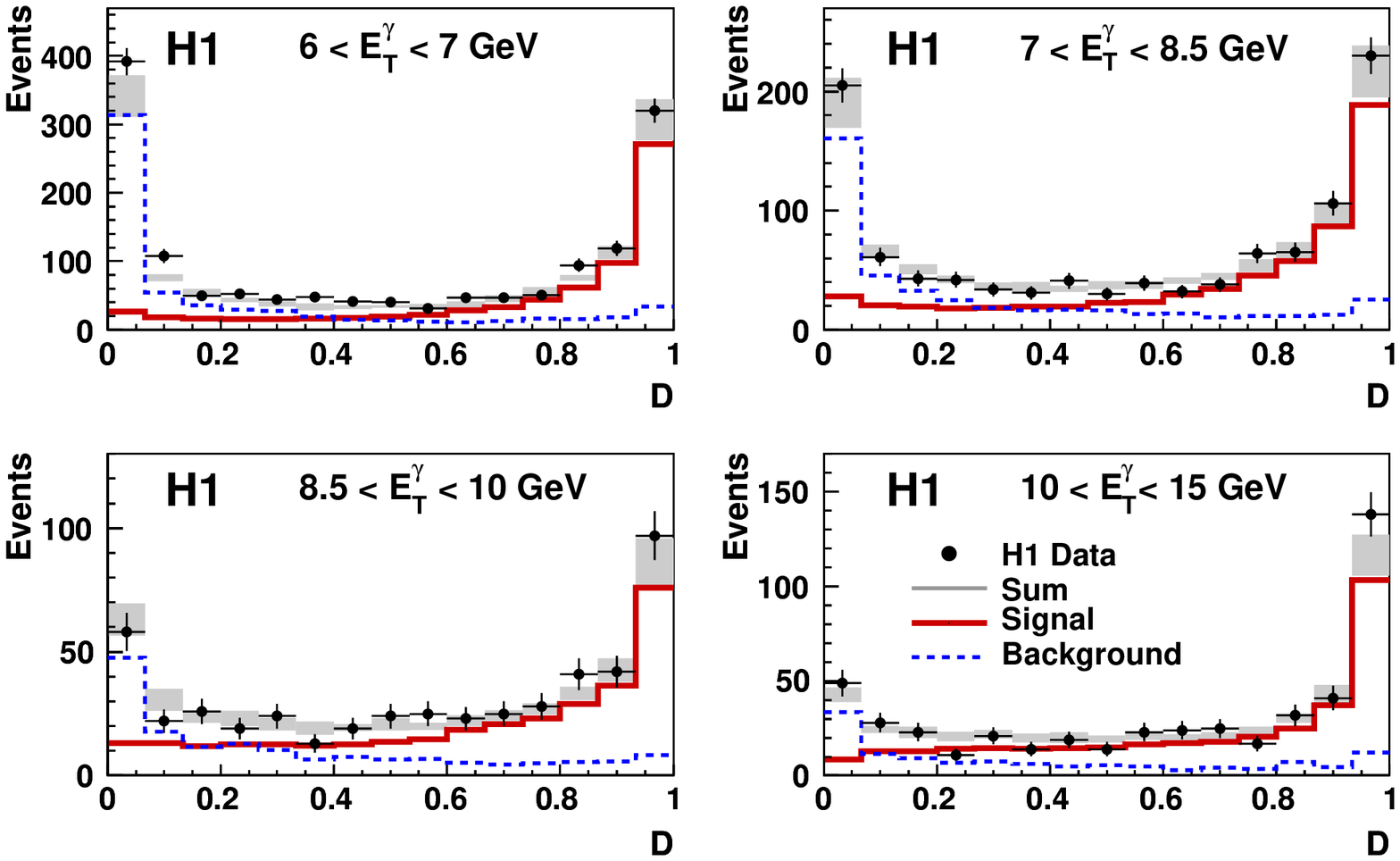}
    \caption{
The distribution of the discriminant $D$ for $0.94<\etag<1.42$ in four different $\etg$ bins. $D$ is used in the identification of
prompt photon candidates for events that have passed the event
selection. Data are shown as points with error bars for the statistical error.
The expectation from PYTHIA for the signal (background) 
is shown as a full (dashed) line. The signal and background contributions are scaled by a factor $1.5$ and $1.7$ on average, respectively.
 The shaded 
 band shows the sum of the MC predictions with the systematic uncertainty from the description of the shower shapes.}
    \label{fig:discriminant}
  \end{center}
\end{figure}\vspace*{\fill}

\vspace*{\fill}\begin{figure}[ht]
  \begin{center}
    \includegraphics[width=1.\textwidth]{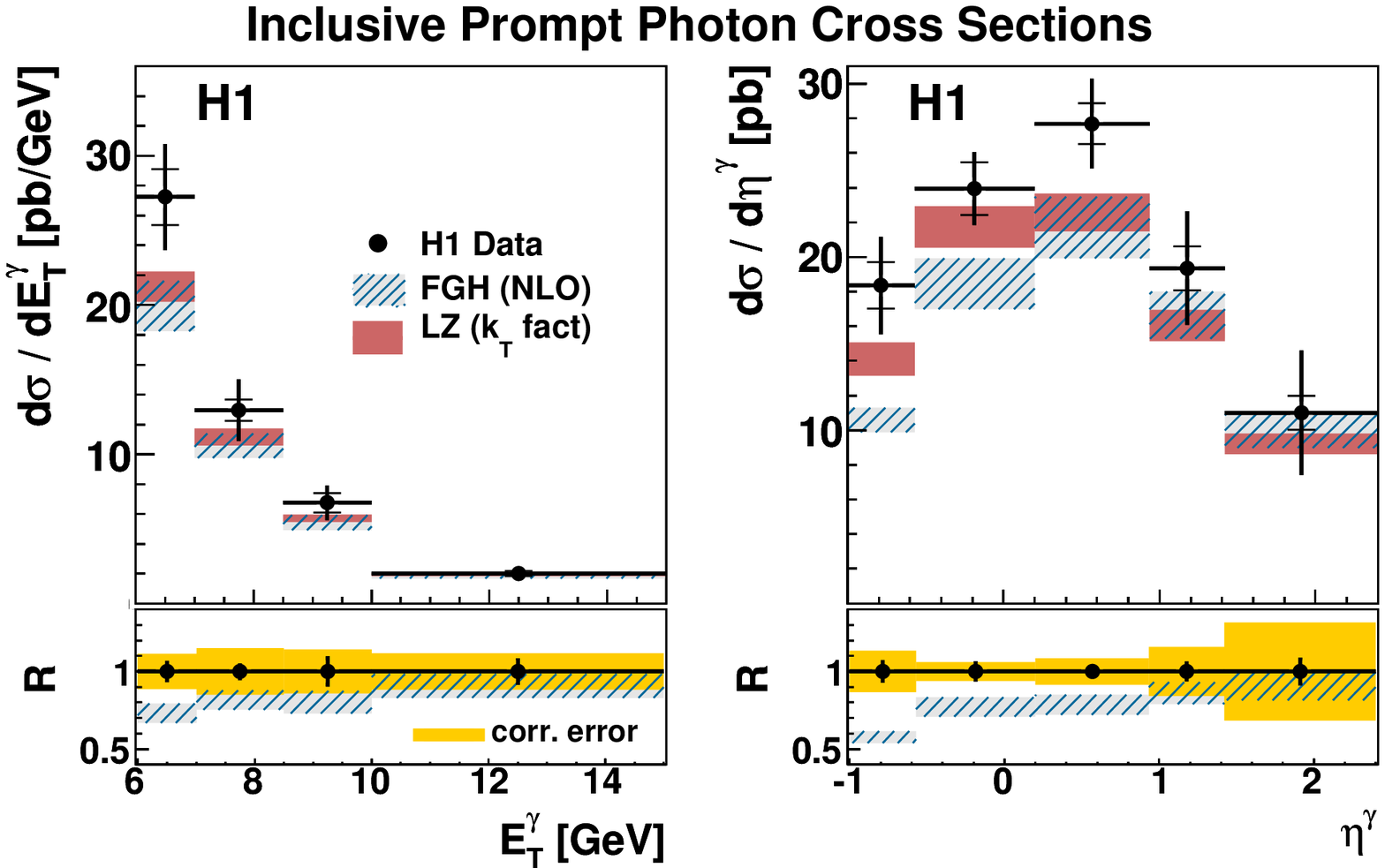}
    \caption{Inclusive differential prompt photon cross sections
     a)~$\mathrm{d}\sigma/\mathrm{d}\etg$ and b)~$\mathrm{d}\sigma/\mathrm{d}\etag$
    in the kinematic range specified in table~\ref{tab:kinem}.
    The inner error bars on the data points indicate the uncorrelated error including the statistical error,
    the full error bars contain in addition the correlated errors added in quadrature.
The data are compared to a QCD   calculation based on the collinear factorisation in 
NLO (FGH)~\cite{heinrich,heinrich2} and to a QCD calculation based on the $k_T$ factorisation approach (LZ)~\cite{zotov}.
    For each plot the lower figure shows the ratio 
     of NLO QCD  to the measured cross section
      ($R = \sigma_{\mathrm{ FGH}}/\sigma$) as a the hatched band.
       The width of this band shows the uncertainty from the NLO
       calculation only.
      The data points are shown at $R=1$ and their bars indicate the
      experimental uncorrelated uncertainty.
       The correlated experimental uncertainty of the data is indicated by the
      shaded area.
   }

    \label{fig:xsec_incl}
  \end{center}
  \begin{picture}(0,0)
  \setlength{\unitlength}{1.0cm}
  \put(7.,14.2){\textsf{a)}}
  \put(15.,14.2){\textsf{b)}}
  \end{picture}
\end{figure}\vspace*{\fill}

\vspace*{\fill}\begin{figure}[ht]
  \begin{center}
 \includegraphics[width=0.97\textwidth]{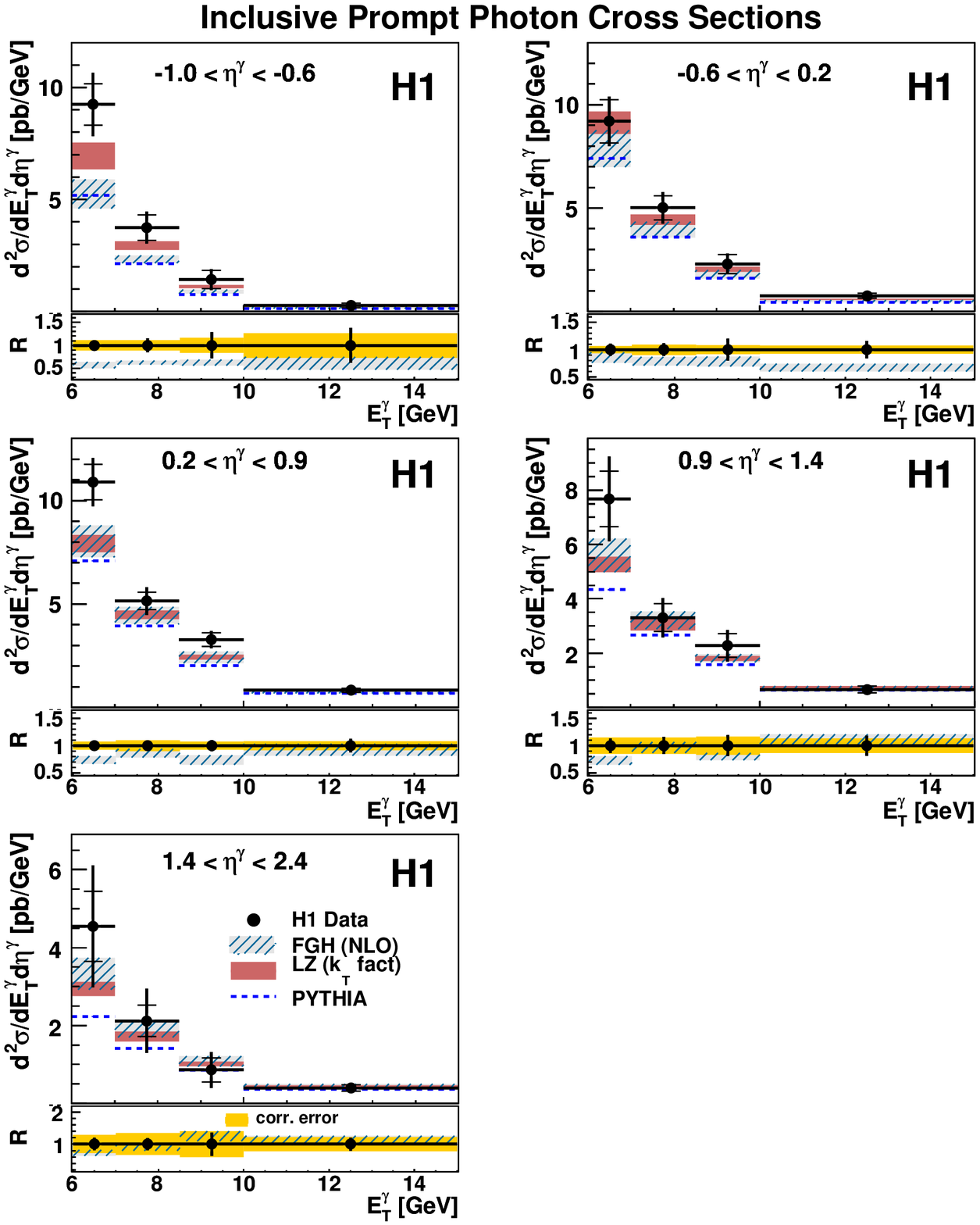}
  \caption{Inclusive double differential prompt photon cross sections
    $\mathrm{d}^2\sigma/\mathrm{d}\etg \mathrm{d}\etag$ for five bins in $\etag$. 
The errors between different $\etag$ bins are correlated.
    The kinematic range is specified in table~\ref{tab:kinem}.
    The cross sections are compared to theoretical
    calculations  (see caption figure~\ref{fig:xsec_incl}). In addition  the prediction from PYTHIA~\cite{PYTHIA} is shown as dashed line.  }
  \label{fig:xsec_incl2}
\end{center}
\end{figure}\vspace*{\fill}

\vspace*{\fill}\begin{figure}[ht]
  \begin{center}
\includegraphics[width=1.\textwidth]{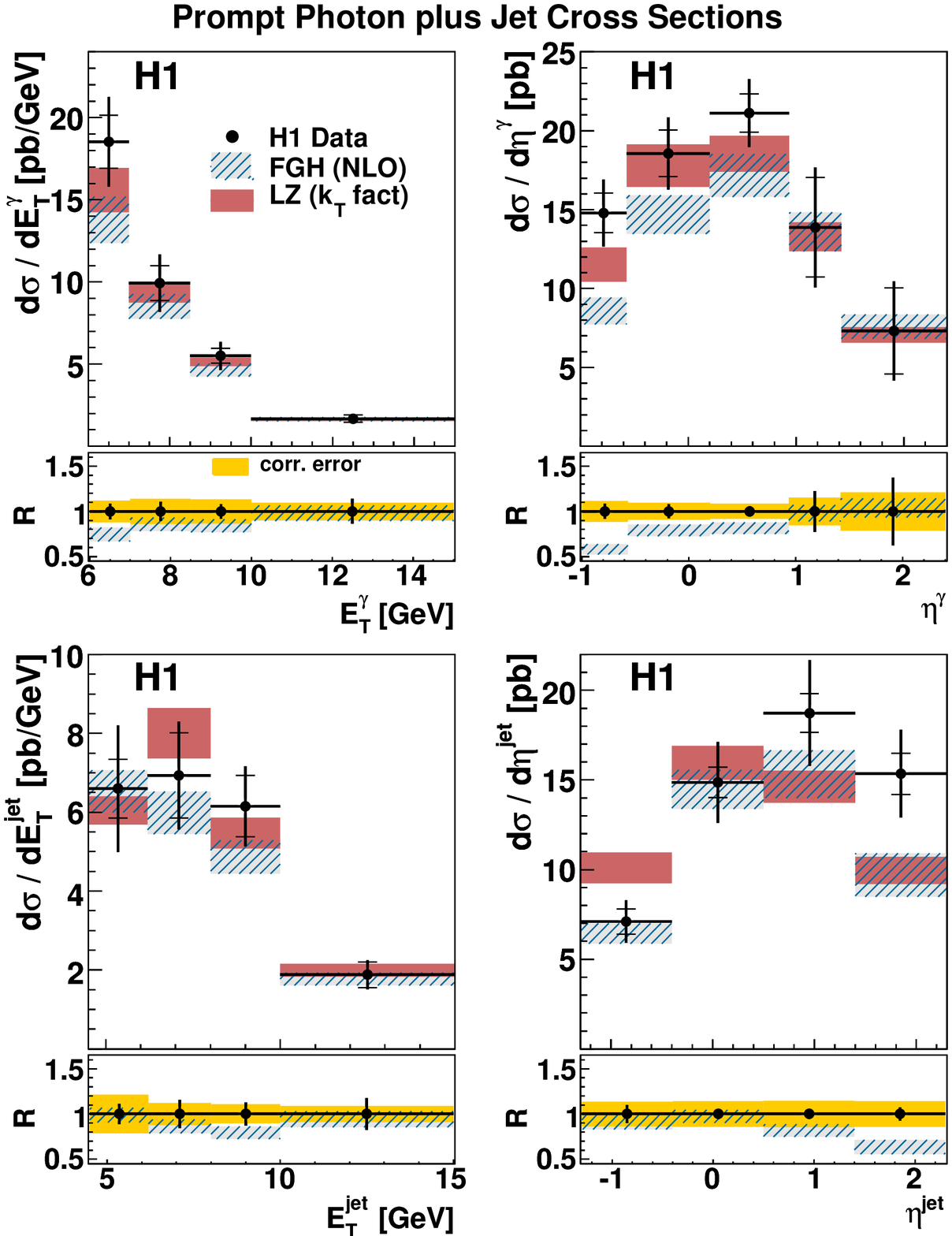}
    \caption{Differential prompt photon plus jet cross sections a)~$\mathrm{d}\sigma/\mathrm{d}\etg$, b)~$\mathrm{d}\sigma/\mathrm{d}\etag$,
   c)~$\mathrm{d}\sigma/\mathrm{d}\etj$ and d)~$\mathrm{d}\sigma/\mathrm{d}\etaj$ 
  in the kinematic range specified in table~\ref{tab:kinem}. The cross sections are compared to theoretical
calculations   (see caption figure~\ref{fig:xsec_incl}).}
    \label{fig:xsec_excl}
  \end{center}
  \begin{picture}(0,0)
  \put(68,214.5){\textsf{a)}}
  \put(148,214.5){\textsf{b)}}
  \put(68,117){\textsf{c)}}
  \put(148,117){\textsf{d)}}
  \end{picture}
\end{figure}\vspace*{\fill}
\vspace*{\fill}\begin{figure}[ht]
  \begin{center}

    \includegraphics[width=1.\textwidth]{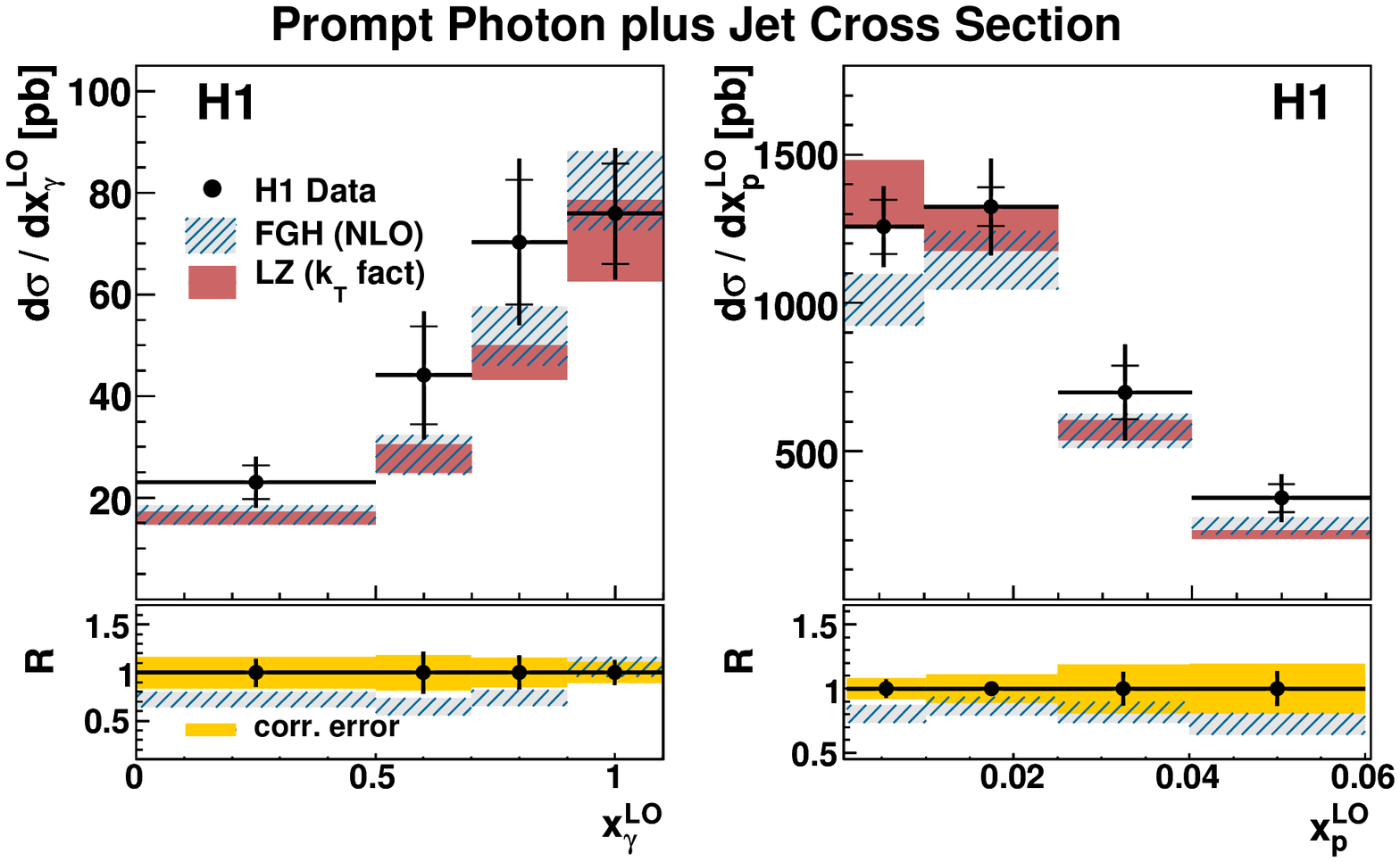}
    \caption{Differential prompt photon plus jet cross sections a)~$\mathrm{d}\sigma/\mathrm{d}\xglo$ and b)~$\mathrm{d}\sigma/\mathrm{d}\xplo$,
   in the kinematic range specified in table~\ref{tab:kinem}. 
  The cross sections are compared to theoretical
calculations   (see caption figure~\ref{fig:xsec_incl}).    }
    \label{fig:xsec_xg}
  \end{center}
  \begin{picture}(0,0)
  \setlength{\unitlength}{1.0cm}
  \put(7.,11.3){\textsf{a)}}
  \put(10.2,11.3){\textsf{b)}}
  \end{picture}
\end{figure}\vspace*{\fill}
\vspace*{\fill}\begin{figure}[ht]
  \begin{center}
\includegraphics[width=0.97\textwidth]{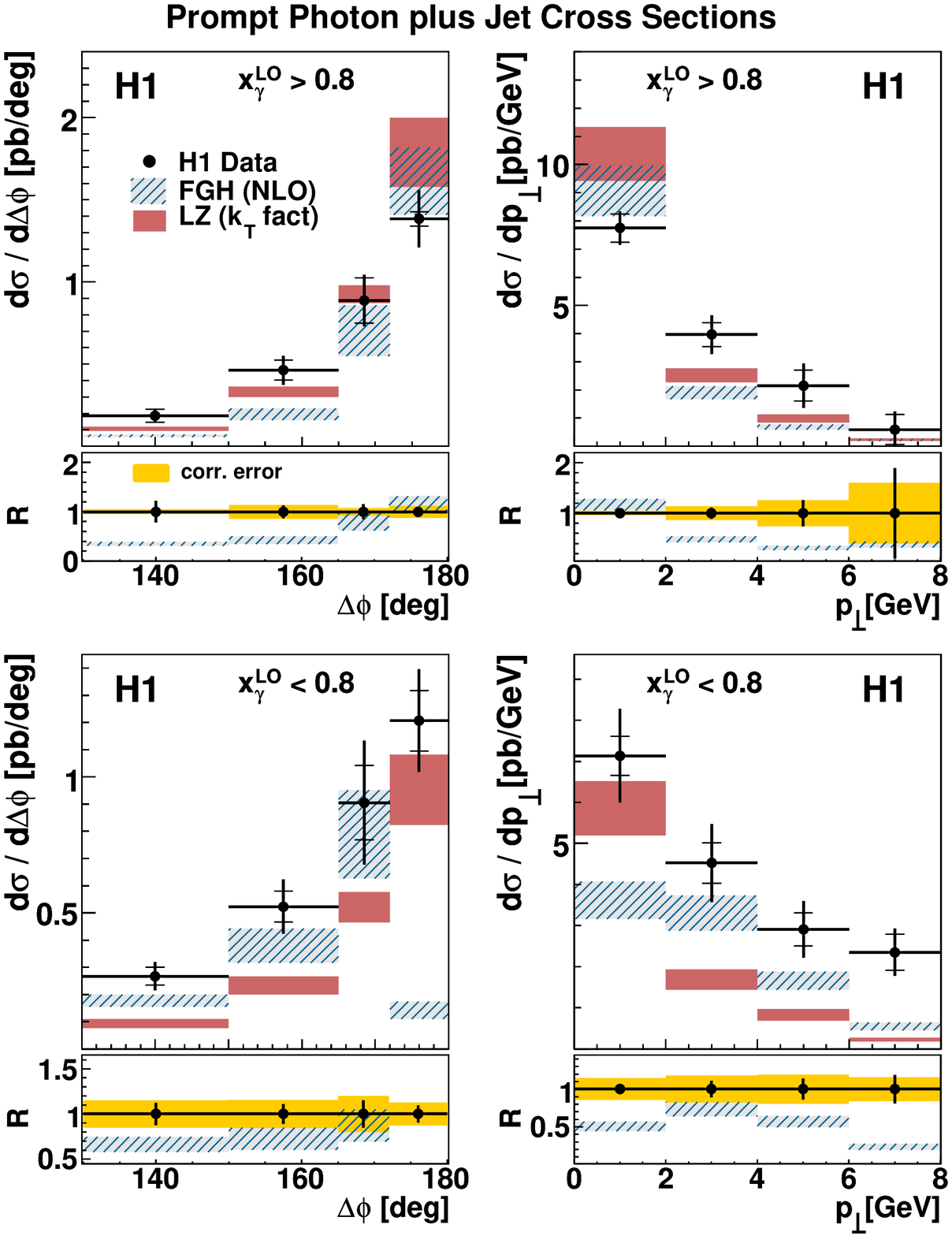}
    \caption{
Differential  prompt photon plus jet cross sections 
 $\mathrm{d}\sigma/\mathrm{d}\Dphi$ and $\mathrm{d}\sigma/\mathrm{d}\pperp$ the photon momentum transverse to the jet direction,  the difference in azimuthal angle between the photon and the jet.
 The kinematic range is specified in table~\ref{tab:kinem}. 
Figures a) and b) show the cross section for $\xglo>0.8$, c) and d) for  $\xglo<0.8$.
The cross sections are compared to theoretical
calculations  (see caption figure~\ref{fig:xsec_incl}).}
    \label{fig:xsec_pperp}
  \end{center}
  \begin{picture}(0,0)
  \put(70,220){\textsf{a)}}
  \put(100,220){\textsf{b)}}
  \put(65,124){\textsf{c)}}
  \put(100,124){\textsf{d)}}
  \end{picture}
\end{figure}\vspace*{\fill}
\end{document}